\documentclass[11pt,a4paper]{article}
\usepackage[T1]{fontenc}
\usepackage{lmodern}
\usepackage[english]{babel}
\usepackage[margin=25mm,headheight=14pt]{geometry}
\usepackage{amsmath,amssymb,mathtools,mathrsfs,booktabs,array,longtable}
\usepackage{microtype}
\usepackage{tikz}
\usepackage{fancyhdr}
\usepackage[colorlinks=true,linkcolor=blue,citecolor=blue,urlcolor=blue]{hyperref}
\allowdisplaybreaks[2]
\numberwithin{equation}{section}
\newcommand{\dd}{\mathrm d}
\newcommand{\Tr}{\operatorname{Tr}}
\newcommand{\Disc}{\operatorname{Disc}}
\newcommand{\remP}{\operatorname{rem}}
\newcommand{\G}{\mathcal G}
\newcommand{\cyc}{\mathop{\sum}\limits_{\mathrm{cyc}}}
\newcolumntype{R}[1]{>{\raggedright\arraybackslash}p{#1}}
\title{Projective symmetries and differential recursions\\
for the large-$N$ Gross--Witten--Wadia\\and one-dimensional QCD matrix models}
\author{S.~Voloshyn\\
\small Bogolyubov Institute for Theoretical Physics, National Academy of Sciences of Ukraine, Kyiv, Ukraine}
\begin{document}
\maketitle
\begin{abstract}
We compare the fixed-charge planar Gross--Witten--Wadia (GWW)
model with source-free one-dimensional QCD at large $N$ and
$N_f/N$ fixed. In both models the resolvent and one-cut condition
give an algebraic equation for a derivative of the free energy:
a quartic for GWW and a sextic for QCD. Finite-$N$ Toeplitz
determinants independently check the selected QCD branch.
Signed permutations of homogeneous parameter triples realize $S_4$;
the resulting action on the algebraic branches and operator labels
gives a common projective description of the two models. For QCD,
the sextic yields a differential recurrence linking small-charge
rows two apart. The same second-order PDE gives a finite-width
column recursion that reconstructs the coefficient pyramid from
its GWW boundary. A distinct boundary, obtained from the first
resummed coefficient, is the Narayana triangle. Its deformation
coefficients have leading poles determined by the critical GWW series.
Exact polynomial certificates establish nonnegativity through the
sixth deformation order for every row. A formally
collapsed slice counts decorated bipartite plane trees and admits
an exact symmetric product formula. The many-flavor scaling
reduces the sextic to the quartic and retains transverse
projective data: the QCD divided-difference triple coalesces
at this boundary, whereas the limiting quartic satisfies the full
standard GWW third-order system. We distinguish algebraic degree,
mixed differential order, and fixed-parameter ordinary differential
order throughout.
\end{abstract}

\section{Introduction}\label{sec:intro}

Large-$N$ unitary matrix integrals often admit several descriptions that
appear quite different when expressed in their original variables. A
small-source expansion organizes the answer into rational coefficient
functions. A spectral construction gives an algebraic relation for a
mass or source derivative of the free energy. Differential recursions
relate successive expansion coefficients, whereas functional
transformations connect expressions obtained in different parameter
regions. The purpose of the present account is to place these structures
in a common framework for two closely related models: the
Gross--Witten--Wadia (GWW) model~\cite{GrossWitten1980,Wadia1980} in a
fixed determinant-charge sector and
the source-free one-dimensional QCD model in the joint large-$N$,
large-$N_f$ limit (for source-deformed Polyakov-loop extensions,
see Refs.~\cite{VoloshynUN2025,VoloshynSUN2025}).

We organize the comparison in fixed determinant-charge sectors,
represented by $U(N)$ integrals. The $SU(N)$ partition functions are
reconstructed by summing over these sectors. Earlier large-$N$
results for the $SU(N)$ GWW integral and the one-dimensional QCD
partition function are given in
Refs.~\cite{BCV2020,BCVY2024QCD}, respectively.

Determinant deformations of GWW provide a related line of work.
Russo analyzed the planar phases of fermion-inspired $U(N)$
deformations~\cite{Russo2020QCD2}; Russo and Tierz studied
characteristic-polynomial insertions using Toeplitz determinants
and Fisher--Hartwig asymptotics~\cite{RussoTierz2020}. Santilli and
Tierz established exact relations between determinant-deformed
unitary ensembles and other random matrix ensembles and showed that
their large-$N$ phase diagrams can differ~\cite{SantilliTierz2020}.
Malik studied the spectral density and phase structure of
QCD-inspired $U(N)$ and $SU(N)$ models with linear sources and
determinant terms~\cite{Malik2026QCDInspired}. Malik and Ahmed
examined a complex unitary potential with Fisher--Hartwig
singularities~\cite{MalikAhmed2026}. The comparison here centers on
the algebraic generating functions at fixed charge, their projective
symmetries, and the differential recursions of their coefficients.

The algebraic curves organize the paper. Sections~\ref{sec:gww}
and \ref{sec:qcdcurve} start from the planar resolvent and the
one-cut condition, obtaining the quartic and the sextic. The
projective coordinates of Section~\ref{sec:models} make their
parameter symmetries explicit, while the coefficient identities
and generating PDEs are established on the curves.
Section~\ref{sec:qcddiff} uses the QCD sextic to derive the
small-charge recurrence, its generating equation, and the
pyramid with GWW and Narayana boundaries.
Section~\ref{sec:limit} then explains how the GWW boundary also
arises as a scaling limit of the QCD integral and curve.
The finite-$N$ calculation in Section~\ref{sec:qcd-toeplitz-check}
tests the chosen branch independently of these derivations.

The central observation is that the relevant parameter transformations
become elementary when the parameters are represented by homogeneous
triples. For GWW the triple is $[u:h:i]$, where $u=q/N$ and $h$ is the
source. For one-dimensional QCD it is $[k:k+1:ku]$, where $k=N_f/N$ and
$u=q/N_f$. Permutations and sign changes of these coordinates induce
rational, and sometimes complex, transformations in the usual affine
parameters. Their apparent complexity is largely a consequence of
restoring an affine normalization after performing a simple homogeneous
operation.

Geometrically, the quartic and sextic describe algebraic coverings
of the corresponding parameter charts. The signed permutations
act on the base coordinates and transport roots between fibers;
in QCD the mass is a separate coordinate. For GWW a permutation
can change the sign of the orientation covariant, while the
normalized QCD root $W$ is invariant on its transported sheet.
The physical germ is selected by boundary data of the integral.
This distinction is why projective equivalence alone does not
assert equality of separately selected thermodynamic branches.

This projective viewpoint also clarifies what it means for a
differential equation to respect a symmetry. A group transformation can
permute several component operators even when a particular combination
of them is preserved up to a scalar factor. Conversely, a covariant
transformation law for one complete operator does not determine how all
possible decompositions of that operator behave. The distinction is
visible already in GWW. There, the difference of two operators can
change sign under a duality transformation, preserving its homogeneous
equation, while the two component operators are exchanged. It is
therefore natural to seek a comparable formulation in one-dimensional
QCD rather than infer the operator structure solely from an invariant
partial differential equation.

A second issue concerns differential order. The algebraic degree of a
generating function, the total order of a partial differential equation,
and the order of an ordinary differential equation obtained by fixing
external parameters are separate quantities. In particular, an
algebraic function of degree six can satisfy a mixed partial
differential equation of order two without admitting a second-order
ordinary differential equation in the mass alone. Similarly, replacing
a free energy by a suitable first derivative can lower the order of its
partial differential equation, whereas an invertible change of
coordinates preserves the order of a given differential operator.
These distinctions are essential when comparing the explicit
third-derivative GWW raising recursion with the second-order
parameter constraints that arise in QCD.

We retain the intermediate normalizations needed to move between the
free energy, its algebraic derivative, and its projective generating
function. This is particularly useful for interpreting earlier GWW
recursions, whose generating function differs from the regular free
energy by a sign and an elementary quadratic term. For QCD, a specific
logarithmic subtraction converts an affine transformation of the free
energy into a homogeneous transformation law. In those variables, the
permuted operator triple has a short expression in Euler derivatives.

The presentation concerns analytic fixed-sector branches before any
sector summation or charge extremization. Polynomial identities
are established at the level of the algebraic curve, and therefore hold
for every regular local root of the relevant polynomial. A physical
branch requires additional information from the matrix integral,
including its small-source boundary condition and the continuation of
the saddle contour. These qualifications enter at the points where
they affect the interpretation of a formula. They do not restrict the
purely algebraic or operator identities derived below.

The comparison is organized around the following results.
\begin{enumerate}
\item The one-cut resolvent gives the GWW quartic and QCD sextic
(Sections~\ref{sec:gww} and \ref{sec:qcdcurve}). The GWW raising
recursion is the reference case; the Toeplitz calculation is an
independent test of the QCD branch.
\item The projective $S_4$ maps act on both curves. The induced
operator action is a literal permutation in homogeneous coordinates
and involves normalization factors in the physical charts.
\item The QCD row recurrence \eqref{eq:pyr-row-rec} and generating
PDE \eqref{eq:pyr-generating} yield the explicit column
reconstruction \eqref{eq:pyr-column-rec}. The first column
is GWW, and the leading resummed boundary is Narayana at all
orders. The collapsed slice admits a separate exact tree count.
\item The many-flavor scaling relates the matrix integrals, the
curves and the coefficient expansions (Section~\ref{sec:limit}).
The limit needs transverse coordinates and does not identify the
particular operator triples component by component.
\end{enumerate}

\section{Sector integrals, normalizations, and projective coordinates}
\label{sec:models}

\subsection{Fixed-charge integrals and algebraic derivatives}

We use the GWW sector integral
\begin{equation}\label{eq:gint}
 D_q^{\mathrm G}(N,h)=\int_{U(N)}\dd U\,(\det U)^q
 \exp\left[\frac{Nh}{2}\Tr(U+U^{-1})\right],\qquad u=\frac qN.
\end{equation}
The symmetry $D_{-q}^{\mathrm G}=D_q^{\mathrm G}$ allows us to choose
$u\geq0$ on the real fixed-sector branch. For algebraic continuation,
$u$ will be treated as a complex parameter. The planar free energy and
its source derivative are
\begin{equation}\label{eq:gnorm}
 \Psi_{\mathrm G}(u,h)=\lim_{N\to\infty}N^{-2}\log D_q^{\mathrm G},
 \qquad g(u,h)=\frac h2\partial_h\Psi_{\mathrm G}
             =x\partial_x\Psi_{\mathrm G},\quad x=h^2.
\end{equation}
The exact finite-$N$ determinant
\begin{equation}
 D_q^{\mathrm G}(N,h)=\det_{1\leq i,j\leq N}I_{q+i-j}(Nh)
\end{equation}
provides an independent definition and fixes the normalization of the
small-source expansion. Its role here is to identify the sector and its
boundary data; the planar algebraic description can be developed without
expanding this determinant in representations.

For source-free one-dimensional QCD we consider
\begin{equation}\label{eq:qint}
 D_q^{\mathrm Q}(N,N_f,h)=\int_{U(N)}\dd U\,(\det U)^q
 \det(1+hU)^{N_f}\det(1+hU^{-1})^{N_f}.
\end{equation}
The two determinants are retained exactly. The parameters in the
Veneziano limit are
\begin{equation}\label{eq:charges}
 k=\kappa=\frac{N_f}{N},\qquad u=\frac q{N_f},\qquad
 v=ku=\frac qN.
\end{equation}
The QCD charge $u$ thus differs from the GWW charge by a factor $k$.
This distinction will be maintained explicitly in the scaling limit.
For positive physical charge the source-free determinant has the
saturation range $0\leq u\leq1$, whereas the GWW sector has no
corresponding upper bound.

The static mass factor has been removed from
\eqref{eq:qint}. With $h=e^{-m}$, the free energy normalized by $NN_f$
and the determinant free energy normalized by $N^2$ are related by
\begin{equation}\label{eq:qnorm}
 \Psi_{\mathrm Q}=\lim N^{-2}\log D_q^{\mathrm Q},\qquad
 S(k,u;h)=-\log h+\frac{\Psi_{\mathrm Q}(k,u;h)}k.
\end{equation}
We define
\begin{equation}\label{eq:qGR}
 \G_{\mathrm Q}=x\partial_x\Psi_{\mathrm Q},\qquad
 R=\frac{\G_{\mathrm Q}}k-\frac12=\frac h2S_h.
\end{equation}
The additive $1/2$ is a normalization effect of the static mass factor.
For example, $\partial_mS=-2R$ in the convention of
\eqref{eq:qnorm}. Selecting a sector or combining several sectors is
a subsequent operation.

In what follows, the term algebraic generating function refers to
$g$, $\G_{\mathrm Q}$, $R$, or an explicitly normalized version of
these derivatives. The free energies themselves contain logarithms.
Their algebraic derivatives determine them up to functions of the
external parameters, which must be fixed by the sector normalization.
This integration freedom is also responsible for additive terms in the
functional transformations of the free energy.

\subsection{Homogeneous triples and the finite projective group}

A homogeneous triple represents the same projective point after a
common nonzero rescaling,
\begin{equation}
 [a:b:c]=[\lambda a:\lambda b:\lambda c].
\end{equation}
In the chart $a=1$, let $p=b/a$ and $q=c/a$. The three elementary
permutations of the homogeneous coordinates act as
\begin{equation}\label{eq:chartmaps}
 (p,q)\xrightarrow{bc}(q,p),\qquad
 (p,q)\xrightarrow{ab}(1/p,q/p),\qquad
 (p,q)\xrightarrow{ac}(p/q,1/q).
\end{equation}
The denominators in these transformations arise from the restoration of
the chart. They are the elementary origin of the fractional-linear
relations encountered in coefficient expansions.

The two models admit the triples
\begin{equation}\label{eq:triples}
 [a:b:c]_{\mathrm G}=[u:h:i],\qquad
 [a:b:c]_{\mathrm Q}=[k:k+1:ku].
\end{equation}
Their different dependence on $h$ is significant. The GWW source is a
coordinate of the triple and can be exchanged with the charge by a
projective transformation. In QCD the mass remains an additional
variable, while the triple consists of flavor and charge data.

Together with independent sign changes, permutations generate
\begin{equation}\label{eq:group}
 \Gamma=\bigl((\mathbb Z_2)^3/\mathbb Z_2\bigr)\rtimes S_3
 \simeq(\mathbb Z_2)^2\rtimes S_3\simeq S_4.
\end{equation}
The quotient removes the common sign, which is a projective rescaling.
The resulting group has order $24$. On the three squared coordinates
$A_1=a^2$, $A_2=b^2$, and $A_3=c^2$, sign changes act trivially, and
only the quotient $S_3$ permutes the coordinate labels. This is the
permutation action that will exchange the differential operators.

This finite action should be distinguished from invariance under the
entire projective linear group. It should also be distinguished from a
Galois group over a fixed coefficient field: the transformations in
\eqref{eq:group} change external parameters, whereas a Galois action
fixes the prescribed coefficient field. A relation between these two
structures would require a separate analysis.

\subsection{Symmetric invariants and orientation}

The elementary symmetric polynomials in $A_i$ are
\begin{equation}
 e_1=A_1+A_2+A_3,\qquad
 e_2=A_1A_2+A_1A_3+A_2A_3,\qquad e_3=A_1A_2A_3.
\end{equation}
On $e_1\neq0$ we use the rational projective invariants
\begin{equation}\label{eq:rs}
 r=\frac{e_2}{e_1^2},\qquad s=\frac{e_3}{e_1^3}.
\end{equation}
Indeed, $A_i/e_1$ are the roots of
\begin{equation}
 \lambda^3-\lambda^2+r\lambda-s=0.
\end{equation}
The normalized orientation covariant and its square are
\begin{equation}\label{eq:disc}
 d=\frac{(A_1-A_3)(A_2-A_3)(A_1-A_2)}{e_1^3},\qquad
 d^2=\Delta=r^2-4r^3-4s+18rs-27s^2.
\end{equation}
An odd permutation reverses the sign of $d$, while $r,s$, and
$\Delta$ remain unchanged. This orientation information is needed in
the depressed GWW quartic. The QCD sextic, in contrast, can be written
entirely in the symmetric invariants. Passing from the ordered
coordinates to $r,s$ is locally invertible away from the corresponding
ramification loci, but it identifies points related by permutations.
Consequently an invariant description can conceal an operator
permutation that is transparent in homogeneous coordinates.

\section{The GWW algebraic generating function and its differential system}
\label{sec:gww}

\subsection{A rational coordinate for the fixed-charge branch}

The GWW quartic, its projective transformations and the associated
hierarchy of differential recursions are discussed in
Ref.~\cite{Voloshyn2026Hierarchies}. Here we give a short resolvent
derivation in the same form as the QCD calculation in
Section~\ref{sec:qcd-spectral-origin}: a rational spectral function,
a one-cut double-zero condition, and elimination to an algebraic
relation for the free-energy derivative.

The algebraic curve can be obtained from a one-cut planar construction.
For nonzero charge, the saddle contour generally moves into the complex
eigenvalue plane. Combining the Haar measure with the determinant
insertion gives the effective potential
\begin{equation}
 V(z)=-\frac h2\left(z+\frac1z\right)-(u-1)\log z.
\end{equation}
Let $\omega(z)$ be the planar resolvent, normalized by
$\omega(z)=z^{-1}+m_1z^{-2}+O(z^{-3})$ at infinity. For a single cut
with endpoints $a_*,b_*$, set $\rho^2=a_*b_*$ and choose the square
root $\mathcal R(z)=\sqrt{(z-a_*)(z-b_*)}$ so that
$\mathcal R(z)\sim z$ and $\mathcal R(0)=-\rho$. Matching the
second-order pole at the origin and the constant at infinity gives
\begin{equation}\label{eq:gcurve}
 V'(z)-2\omega(z)=-\frac{h}{2z^2}\left(z+\frac1\rho\right)
 \mathcal R(z).
\end{equation}
If $S_*=a_*+b_*$, the matching conditions are
\begin{equation}
 \frac h2\left(\rho-\frac{S_*}{2\rho^2}\right)=1-u,
 \qquad S_* =\frac2\rho-\frac{4(1+u)}h.
\end{equation}
Eliminating $S_*$ produces the rational parameterization
\begin{equation}\label{eq:gparam}
 h=\frac{2\rho[1+u+(u-1)\rho^2]}{1-\rho^4},\qquad
 g=\frac h4\left(\rho+\frac1\rho\right)
       -\frac{1+\rho^4}{2(1+\rho^2)^2}.
\end{equation}
The second formula follows from the first moment of the resolvent and
the sector identity $m_1-m_{-1}=-2u/h$. The chosen small-source branch
has $\rho\sim h/[2(1+u)]$.

For completeness, define the moments by
$m_j=\int\rho(\xi)\xi^j\,\dd\xi$. Planar factorization gives
\begin{equation}\label{eq:gww-loop-equation}
 \omega(z)^2-V'(z)\omega(z)+f_{\mathrm G}(z)=0,\qquad
 f_{\mathrm G}(z)=-\frac{h}{2z}-\frac{hm_{-1}}{2z^2}.
\end{equation}
Indeed, inserting the rational force
$V'=-h/2+(1-u)/z+h/(2z^2)$ into
$f_{\mathrm G}=\int[V'(z)-V'(\xi)]\rho(\xi)/(z-\xi)\,\dd\xi$
and matching the $1/z$ term at infinity yields this expression.
The source derivative and the scaling Ward identity give
\begin{equation}
 g=\frac h4(m_1+m_{-1}),\qquad m_1-m_{-1}=-\frac{2u}{h},
 \qquad hm_{-1}=2g+u.
\end{equation}
Thus the squared spectral function has a quartic numerator:
\begin{align}\label{eq:gww-spectral-quartic}
 (V'-2\omega)^2&=\frac{Q_{\mathrm G}(z)}{4z^4},\\
 Q_{\mathrm G}(z)&=[-hz^2+2(1-u)z+h]^2
                    +8hz^3+8(2g+u)z^2.\notag
\end{align}
The one-cut ansatz \eqref{eq:gcurve} makes this numerator
$h^2(z+\rho^{-1})^2(z-a_*)(z-b_*)$; hence it has a double zero.
More precisely, direct elimination gives the exact identity
\begin{equation}\label{eq:gww-disc-origin}
 \boxed{\operatorname{Disc}_{z}Q_{\mathrm G}
                      =2^{20}h^4P_4(g;u,h).}
\end{equation}
The polynomial $P_4$ is displayed in \eqref{eq:P4} below.
This supplies the same spectral-discriminant construction that gives
the QCD sextic. Its physical local branch is fixed by
$g=u/2+h^2/[4(1+u)]+O(h^4)$; the discriminant equation alone does
not select that branch.

For $u,h>0$, the real parameter $0<\rho<1$ is particularly useful:
\begin{equation}
 \left.\frac{\partial h}{\partial\rho}\right|_u
 =\frac{2\{u(1+\rho^2)^3+(1-\rho^2)^3\}}{(1-\rho^4)^2}>0.
\end{equation}
It therefore labels a unique solution within this one-cut family for
every positive pair $(u,h)$. The algebraic discussion below concerns
that family and its analytic continuations. A comparison with all other
possible complex-contour saddles would be needed to establish a global
dominance statement.

\subsection{The quartic and its projective form}

Elimination of $\rho$ in \eqref{eq:gparam} gives the monic quartic
\begin{align}\label{eq:P4}
0=P_4(g;u,h)={}&g^4+\frac{1+u^2-h^2}{2}g^3\notag\\
&+\frac{(1-u^2)^2+2(u^2-4)h^2+h^4}{16}g^2\notag\\
&+\frac{-4u^2(1+u^2)+(u^2-1)h^2+5h^4}{32}g\notag\\
&-\frac{u^2(1+u^2)^2}{64}
 -\frac{u^2(3u^2-5)h^2}{64}
 +\frac{(1-12u^2)h^4}{256}-\frac{h^6}{64}.
\end{align}
The boundary condition
\begin{equation}\label{eq:gsmall}
 g=\frac u2+\frac{h^2}{4(1+u)}+O(h^4)
\end{equation}
identifies the desired local root. Both the polynomial and the boundary
condition are needed: the polynomial alone specifies all four branches.

For $[a:b:c]=[u:h:i]$, the symmetric polynomials become
\begin{equation}
 e_1=u^2+h^2-1,\qquad e_2=u^2h^2-u^2-h^2,\qquad e_3=-u^2h^2.
\end{equation}
Introduce the centered variable and the unnormalized orientation factor
\begin{equation}\label{eq:tau}
 \tau=-4g-\frac{1+u^2-h^2}{2},\qquad
 \delta=(1+u^2)(1+h^2)(u^2-h^2).
\end{equation}
The quartic then takes the homogeneous form
\begin{align}\label{eq:homP4}
0={}&\tau^4+\frac{12e_2-e_1^2}{2}\tau^2+4\delta\tau\notag\\
 &+\frac{e_1^4}{16}+\frac{e_1^2e_2}{2}-3e_2^2+18e_1e_3.
\end{align}
Every term has homogeneous degree eight when $\tau$ has degree two
in $a,b,c$. Substitution of \eqref{eq:tau} into the left-hand side
of \eqref{eq:homP4} gives exactly $256P_4$. The projective form
therefore preserves the complete algebraic relation, including its
normalization.

Under an odd permutation, $\delta$ and $\tau$ both change sign.
After restoring $c=i$, the elementary transpositions give
\begin{equation}\label{eq:gtransforms}
\begin{array}{c|c|c}
 & (u',h')&\tau'\\\hline
 ab&(h,u)&-\tau\\
 ac&(-1/u,ih/u)&\tau/u^2\\
 bc&(iu/h,-1/h)&\tau/h^2
\end{array}
\end{equation}
These are pure permutations of the homogeneous coordinates followed by
rescaling. Additional sign changes give alternative representatives,
such as $(1/u,ih/u)$. In terms of the original generating function,
\begin{equation}\label{eq:grootmaps}
 g'_{ab}=-g-\frac14,\qquad
 g'_{ac}=\frac{g-h^2/4}{u^2},\qquad
 g'_{bc}=\frac{g+(1+u^2-h^2)/4}{h^2}.
\end{equation}
The first relation already shows why an algebraic symmetry need not
preserve a physical root: it maps a positive real $g$ to a negative
one. Equations \eqref{eq:gtransforms} and \eqref{eq:grootmaps} should
therefore be interpreted with the root transported along the parameter
transformation, rather than reselected independently at the image point.

Dividing by the appropriate homogeneous powers yields a description in
projective invariants. Set $w=\tau/e_1$ and $d=\delta/e_1^3$. Then
\begin{equation}\label{eq:orientedP4}
 w^4+\frac{12r-1}{2}w^2+4dw+
 \frac1{16}+\frac r2-3r^2+18s=0.
\end{equation}
The scalar $w$ is odd under a transposition, but $\zeta=dw$ is
invariant. Its polynomial is
\begin{align}\label{eq:invariantP4}
0={}&\zeta^4+\frac{12r-1}{2}\Delta\zeta^2+4\Delta^2\zeta\notag\\
 &+\left(\frac1{16}+\frac r2-3r^2+18s\right)\Delta^2.
\end{align}
On $\Delta=0$, this particular invariant variable loses information
about $w$. The oriented equation \eqref{eq:orientedP4}, or the
original quartic, supplies a regular chart when the invariant
normalization degenerates.

\subsection{An explicit coefficient recursion and the regular free energy}

The quartic branches obey the differential identity
\begin{equation}\label{eq:gPDE}
 h\,g_{hhh}=u\,g_{uuu}.
\end{equation}
It is a linear partial differential equation in two parameters, and its
small-source expansion gives a particularly simple raising relation.
Write
\begin{equation}\label{eq:gseries}
 g=\frac u2+\frac{h^2}{4}
 -\sum_{n\geq0}R_n(u)\left(\frac h2\right)^{2n+2},\qquad
 R_0=\frac{u}{1+u}.
\end{equation}
Comparison of the coefficient of $h^{2n}$ in \eqref{eq:gPDE}
yields
\begin{equation}\label{eq:gRec}
 R_n(u)=\frac{u\,\partial_u^3R_{n-1}(u)}{n(n+1)(2n+1)},
 \qquad n\geq1.
\end{equation}
The right-hand side acts only on the known previous coefficient. Thus
\eqref{eq:gRec} is an explicit differential raising formula, a
property that will be useful in assessing the QCD recursions.
The corresponding signed numerator triangle is
OEIS A336110~\cite{OEISGWWTriangle}; see also
Ref.~\cite{Voloshyn2026Hierarchies} for the differential recursion.

The first coefficients are
\begin{align}
R_0&=\frac{u}{1+u},&R_1&=\frac{u}{(1+u)^4},\\
R_2&=-\frac{2u(u-1)}{(1+u)^7},&
R_3&=\frac{u(5u^2-14u+5)}{(1+u)^{10}}.
\end{align}
They give
\begin{align}
g={}&\frac u2+\frac{h^2}{4(1+u)}
 -\frac{uh^4}{16(1+u)^4}
 +\frac{u(u-1)h^6}{32(1+u)^7}\notag\\
 &-\frac{u(5u^2-14u+5)h^8}{256(1+u)^{10}}+O(h^{10}).
\end{align}
The powers of $2$ originate from the source normalization in
\eqref{eq:gint}. Integrating \eqref{eq:gseries} gives
\begin{equation}\label{eq:gfreeSeries}
 \Psi_{\mathrm G}=u\log h+C(u)+\frac{h^2}{4}
 -\sum_{n\geq0}\frac{R_n(u)}{n+1}\left(\frac h2\right)^{2n+2}.
\end{equation}
The function $C(u)$ is fixed by the matrix integral and cannot be
recovered from the mass derivative alone.

To compare with an operator formulation written directly for the regular
free energy, define
\begin{equation}\label{eq:Pg}
 P(u,h)=\frac{h^2}{4}-[\Psi_{\mathrm G}-u\log h-C(u)]
 =\sum_{n\geq0}P_n(u,h),
\end{equation}
\begin{equation}
 P_n=\frac{R_n(u)}{n+1}\left(\frac h2\right)^{2n+2},\qquad
 g=\frac u2+\frac{h^2}{4}-\frac h2P_h.
\end{equation}
For example,
\begin{equation}
 P_0=\frac{uh^2}{4(1+u)},\quad
 P_1=\frac{uh^4}{32(1+u)^4},\quad
 P_2=-\frac{u(u-1)h^6}{96(1+u)^7}.
\end{equation}
The sign and the factor $n+1$ distinguish $P_n$ from the coefficient
functions in the algebraic derivative. Keeping this dictionary explicit
avoids an apparent disagreement between equivalent recursions.

Let
\begin{equation}\label{eq:gOldOps}
 K_+=\partial_h^2(h\partial_h),\qquad R_+=u\partial_u^3,
 \qquad D=u\partial_u+h\partial_h,
\end{equation}
and introduce
\begin{align}\label{eq:Delta0}
 \Delta_0&=D(D-1)(D-2)\notag\\
 &=u^3\partial_u^3+3u^2h\partial_u^2\partial_h
 +3uh^2\partial_u\partial_h^2+h^3\partial_h^3.
\end{align}
The notation $\Delta_0$ separates this operator from the projective
discriminant $\Delta(r,s)$. In particular,
$K_+=h\partial_h^3+2\partial_h^2$; the second term is part of its
definition. Equation \eqref{eq:gRec} is equivalent to
\begin{equation}\label{eq:gPnRec}
 R_+P_n=K_+P_{n+1}
 =\frac{(2n+3)(2n+4)^2}{h^2}P_{n+1}.
\end{equation}
Summation over $n$ gives
\begin{equation}\label{eq:gPPDEs}
 (R_+-K_+)P=-\frac{u}{1+u},\qquad
 (\Delta_0-K_+)P=-\frac{u}{1+u}.
\end{equation}
Consequently $(\Delta_0-R_+)P=0$. Exchanging $R_+$ and $\Delta_0$
changes the sign of this last operator and preserves its homogeneous
equation. This is a simple example of how a nontrivial permutation of
components can coexist with the preservation of the complete equation.

For the convention $[A,B]=AB-BA$, homogeneity gives
\begin{equation}
 [K_+,D]=2K_+,\qquad [R_+,D]=2R_+,\qquad [\Delta_0,D]=0,
\end{equation}
and hence
\begin{equation}
 [\Delta_0-R_+,D]=-2R_+,\qquad
 [\Delta_0-K_+,D]=-2K_+.
\end{equation}
These identities identify the Euler-operator structure of the
recursions. A connection with conformal field theory would require
additional structure beyond these homogeneity relations.

\subsection{A homogeneous third-order operator triple}

The projective nature of the differential system is most transparent
for the centered algebraic function $\tau$. Extend it from the chart
$c=i$ to a homogeneous function of degree two in $a,b,c$, and define
\begin{equation}\label{eq:gO}
 \mathcal O_1=a\partial_a^3,\qquad
 \mathcal O_2=b\partial_b^3,\qquad
 \mathcal O_3=c\partial_c^3.
\end{equation}
The full system is
\begin{equation}\label{eq:gTriple}
 \mathcal O_1\tau=\mathcal O_2\tau=\mathcal O_3\tau.
\end{equation}
Permutations of the homogeneous coordinates permute these operators
literally. A sign change of a coordinate leaves the corresponding
operator unchanged, since the coordinate and its third derivative
supply four sign factors in total.

To recover the affine system, use homogeneity to write
$c\partial_c\tau=(2-D)\tau$. It follows that
\begin{equation}
 c^3\partial_c^3\tau=(2-D)(1-D)(-D)\tau
 =-D(D-1)(D-2)\tau.
\end{equation}
Since $c^2=-1$ on the chosen chart, $\mathcal O_3\tau=\Delta_0\tau$.
The affine polynomial subtracted in \eqref{eq:tau} is annihilated by
all three third-order operators. Thus \eqref{eq:gTriple} becomes
\begin{equation}\label{eq:gthree}
 u g_{uuu}=h g_{hhh}=\Delta_0g.
\end{equation}
Both independent differences have been checked by exact polynomial
reduction modulo \eqref{eq:P4}. The third pairwise difference is
linearly dependent on the first two, as expected:
\begin{equation}
 (\mathcal O_1-\mathcal O_2)+(\mathcal O_2-\mathcal O_3)
 +(\mathcal O_3-\mathcal O_1)=0.
\end{equation}
A permuted triple therefore need not provide three independent
equations. The role of the triple is to exhibit the coordinate symmetry
of the differential system.

\subsection{An additional second-order invariant equation}

The algebraic function also satisfies differential identities of a
form different from the raising recursion. Write
\eqref{eq:orientedP4} as
\begin{equation}
 w^4+\mathfrak a w^2+\mathfrak b w+\mathfrak c=0,
\end{equation}
where
\begin{equation}
 \mathfrak a=\frac{12r-1}{2},\qquad\mathfrak b=4d,\qquad
 \mathfrak c=\frac1{16}+\frac r2-3r^2+18s.
\end{equation}
On a generic chart with $\mathfrak a\mathfrak b\neq0$, introduce
\begin{equation}\label{eq:gCanonicalVars}
 \alpha=\frac{\mathfrak b^2}{\mathfrak a^3},\qquad
 \beta=\frac{\mathfrak c}{\mathfrak a^2},\qquad
 Y_{\mathrm G}=\frac{\mathfrak b w}{\mathfrak a^2}.
\end{equation}
All three quantities are invariant under the signed-permutation action.
They satisfy the canonical quartic
\begin{equation}\label{eq:gCanonical}
 Y_{\mathrm G}^4+\alpha Y_{\mathrm G}^2
 +\alpha^2Y_{\mathrm G}+\alpha^2\beta=0,
\end{equation}
and the second-order partial differential equation
\begin{equation}\label{eq:gSecond}
 4\alpha^2Y_{\alpha\alpha}+3\alpha^2Y_{\alpha\beta}
 +2\alpha\beta Y_{\beta\beta}-2\alpha Y_\alpha+2Y=0,
 \qquad Y=Y_{\mathrm G}.
\end{equation}
This relation can be derived without using the particular GWW
parameterization. For a root $f(A,B,C)$ of
$f^4+Af^2+Bf+C=0$, implicit differentiation gives
$f_{AC}=f_{BB}$, while weighted homogeneity gives
$2Af_A+3Bf_B+4Cf_C=f$. Differentiating the latter relation with
respect to $C$, eliminating $f_{AC}$, and fixing $A=1$ yields
\begin{equation}
 2f_{BB}+3Bf_{BC}+4Cf_{CC}+3f_C=0.
\end{equation}
The substitution $\alpha=B^2$, $\beta=C$, $Y=Bf$ gives
\eqref{eq:gSecond}. An independent polynomial-remainder calculation
verifies the same identity against \eqref{eq:gCanonical}.

The map from $(r,s)$ to $(\alpha,\beta)$ has Jacobian
\begin{equation}
 \det\frac{\partial(\alpha,\beta)}{\partial(r,s)}
 =-\frac{4096(27s-1)^2}{(12r-1)^6},
\end{equation}
and is locally nonsingular generically. Nevertheless,
\eqref{eq:gSecond} is an additional differential identity, rather than
a coordinate transform that lowers the order of \eqref{eq:gPDE}.
The two statements have different content. In particular, the existence
of \eqref{eq:gSecond} does not replace the explicit third-derivative
raising formula \eqref{eq:gRec} by an explicit second-derivative
raising formula.

\subsection{The integrated free energy and its real branch}

The rational coordinate \eqref{eq:gparam} also makes the free-energy
integration elementary. Define $A=1+u+(u-1)\rho^2$. With logarithms
continued consistently along the chosen branch, one obtains
\begin{align}\label{eq:gexactFree}
\Psi_{\mathrm G}={}&u\log\rho+\frac{u^2}{2}\log(1+\rho^2)
 -\frac{u^2+1}{2}\log A+\frac12\log(1-\rho^2)\notag\\
&+\frac{2u}{1-\rho^2}+\frac{u+1}{1+\rho^2}
 -\frac1{(1+\rho^2)^2}+\frac{u^2}{2}\log u-\frac{3u}{2}.
\end{align}
Direct differentiation gives
$\partial_\rho\Psi_{\mathrm G}=(2g/h)\partial_\rho h$.
The integration constant is fixed by the leading term of the finite-$N$
sector integral,
\begin{equation}
 D_q^{\mathrm G}(N,h)=\left(\frac{Nh}{2}\right)^{Nq}
 \prod_{j=0}^{N-1}\frac{j!}{(q+j)!}[1+O(h^2)],\qquad q\geq0.
\end{equation}
Its planar limit gives
\begin{equation}\label{eq:gC}
 C(u)=\frac{3u}{2}-u\log2+\frac{u^2}{2}\log u
 -\frac{(1+u)^2}{2}\log(1+u)
\end{equation}
in \eqref{eq:gfreeSeries}. An elementary primitive without this
parameter-dependent constant would describe a different sector
normalization.

At zero charge the familiar two GWW branches are recovered as boundary
limits of the parameterization:
\begin{equation}
 \Psi_{\mathrm G}(0,h)=
 \begin{cases}
 h^2/4,&0\leq h\leq1,\\
 h-\tfrac12\log h-\tfrac34,&h\geq1.
 \end{cases}
\end{equation}
For $h>1$, the limit is $u\downarrow0$ together with
$\rho\uparrow1$; substitution of $u=0$ at fixed $\rho<1$ retains
only the weak branch. This illustrates how the same algebraic curve can
contain different physical boundary limits.

The quartic discriminant is
\begin{equation}\label{eq:gDisc}
 \Disc_gP_4=\frac{u^2h^2}{2^{16}}
 \left[(1-u^2-h^2)^3-27u^2h^2\right]^3.
\end{equation}
The nontrivial component in the positive quadrant is
$u^{2/3}+h^{2/3}=1$. If $a=u^{2/3}$ on this curve, the colliding
algebraic root and the continued physical root are
\begin{equation}
 g_{\mathrm{coll}}=\frac{1-3a}{4},\qquad
 g_{\mathrm{ph}}=-\frac{(1-a)^2(2a+1)}4
 +\frac12(a^2-a+1)^{3/2}.
\end{equation}
They coincide only at $(u,h)=(0,1)$. Thus the vanishing of the
polynomial discriminant does not identify an additional physical
singularity of the positive-charge branch by itself.

For completeness, the fixed-sector expansion at $u=0$ and $0<h<1$
can be written in terms of
\begin{equation}\label{eq:glinear}
 a(h)=\log\frac{1+\sqrt{1-h^2}}h-\sqrt{1-h^2}
\end{equation}
as
\begin{align}
 \Psi_{\mathrm G}={}&\frac{h^2}{4}-a(h)u
 +\frac{u^2}{2}\log u\notag\\
 &-\left[\frac34+\frac14\log(1-h^2)\right]u^2+O(u^3).
\end{align}
The nonanalytic term $u^2\log u$ is a property of this fixed-sector
boundary expansion; its thermodynamic interpretation requires a
separate branch and sector selection.

\subsection{A conjectural description-tree refinement}
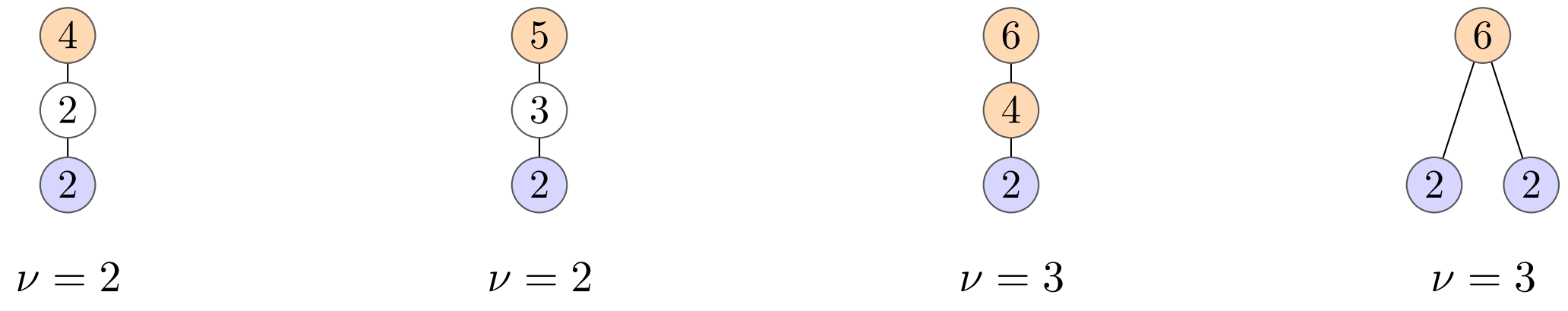
\begin{figure}[htbp]
\centering
\begingroup
\tikzset{descvertex/.style={circle,draw=black!65,fill=white,
 minimum size=4.8mm,inner sep=0pt,font=\small},
 descmax/.style={descvertex,fill=orange!30},
 descleaf/.style={descvertex,fill=blue!16}}
\begin{minipage}[b]{.23\textwidth}\centering
\begin{tikzpicture}[x=.65cm,y=.65cm]
\draw (0,2)--(0,1)--(0,0);
\node[descmax] at (0,2) {$ 4 $};
\node[descvertex] at (0,1) {$ 2 $};
\node[descleaf] at (0,0) {$2$};
\end{tikzpicture}\\[2mm]$\nu=2$
\end{minipage}\hfill
\begin{minipage}[b]{.23\textwidth}\centering
\begin{tikzpicture}[x=.65cm,y=.65cm]
\draw (0,2)--(0,1)--(0,0);
\node[descmax] at (0,2) {$ 5 $};
\node[descvertex] at (0,1) {$ 3 $};
\node[descleaf] at (0,0) {$2$};
\end{tikzpicture}\\[2mm]$\nu=2$
\end{minipage}\hfill
\begin{minipage}[b]{.23\textwidth}\centering
\begin{tikzpicture}[x=.65cm,y=.65cm]
\draw (0,2)--(0,1)--(0,0);
\node[descmax] at (0,2) {$ 6 $};
\node[descmax] at (0,1) {$ 4 $};
\node[descleaf] at (0,0) {$2$};
\end{tikzpicture}\\[2mm]$\nu=3$
\end{minipage}\hfill
\begin{minipage}[b]{.23\textwidth}\centering
\begin{tikzpicture}[x=.65cm,y=.65cm]
\draw (-.65,0)--(0,2)--(.65,0);
\node[descmax] at (0,2) {$6$};
\node[descleaf] at (-.65,0) {$2$};
\node[descleaf] at (.65,0) {$2$};
\end{tikzpicture}\\[2mm]$\nu=3$
\end{minipage}
\endgroup
\caption{All four description trees with two edges.  Blue leaves
and orange maximal internal vertices each contribute one to $\nu$;
white internal vertices do not contribute.  The root is always
maximal.  The distribution $(2,2)$ gives $B_2(z)=2+2z$ and the
signed GWW row $(2,-2)$.}
\label{fig:gww-description-s2}
\end{figure}

\begin{figure}[htbp]
\centering
\begingroup
\tikzset{descvertex/.style={circle,draw=black!65,fill=white,
 minimum size=3.8mm,inner sep=0pt,font=\scriptsize},
 descmax/.style={descvertex,fill=orange!30},
 descleaf/.style={descvertex,fill=blue!16}}
\begin{minipage}[b]{.19\textwidth}\centering
\begin{tikzpicture}[x=.49cm,y=.49cm]
\draw (0,0)--(0,3);
\node[descmax] at (0,3) {$ 4 $};
\node[descvertex] at (0,2) {$ 2 $};
\node[descvertex] at (0,1) {$ 2 $};
\node[descleaf] at (0,0) {$ 2 $};
\draw[->,black!55] (.38,1.5)--(.85,1.5);
\begin{scope}[xshift=.58cm]
\draw[step=1,gray!22,very thin] (0,0) grid (3,3);
\draw[gray!65,dashed,thin] (0,0)--(3,3);
\draw[blue!70!black,very thick] (0,0)--(0,1)--(1,1)--(1,2)--(2,2)--(2,3)--(3,3);
\end{scope}\end{tikzpicture}\\[2mm]
$a=(0,0,0)$
\end{minipage}\hfill
\begin{minipage}[b]{.19\textwidth}\centering
\begin{tikzpicture}[x=.49cm,y=.49cm]
\draw (0,0)--(0,3);
\node[descmax] at (0,3) {$ 5 $};
\node[descvertex] at (0,2) {$ 3 $};
\node[descvertex] at (0,1) {$ 2 $};
\node[descleaf] at (0,0) {$ 2 $};
\draw[->,black!55] (.38,1.5)--(.85,1.5);
\begin{scope}[xshift=.58cm]
\draw[step=1,gray!22,very thin] (0,0) grid (3,3);
\draw[gray!65,dashed,thin] (0,0)--(3,3);
\draw[blue!70!black,very thick] (0,0)--(0,1)--(1,1)--(1,2)--(1,3)--(2,3)--(3,3);
\end{scope}\end{tikzpicture}\\[2mm]
$a=(0,0,1)$
\end{minipage}\hfill
\begin{minipage}[b]{.19\textwidth}\centering
\begin{tikzpicture}[x=.49cm,y=.49cm]
\draw (0,0)--(0,3);
\node[descmax] at (0,3) {$ 4 $};
\node[descvertex] at (0,2) {$ 2 $};
\node[descvertex] at (0,1) {$ 3 $};
\node[descleaf] at (0,0) {$ 2 $};
\draw[->,black!55] (.38,1.5)--(.85,1.5);
\begin{scope}[xshift=.58cm]
\draw[step=1,gray!22,very thin] (0,0) grid (3,3);
\draw[gray!65,dashed,thin] (0,0)--(3,3);
\draw[blue!70!black,very thick] (0,0)--(0,1)--(0,2)--(1,2)--(2,2)--(2,3)--(3,3);
\end{scope}\end{tikzpicture}\\[2mm]
$a=(0,1,0)$
\end{minipage}\hfill
\begin{minipage}[b]{.19\textwidth}\centering
\begin{tikzpicture}[x=.49cm,y=.49cm]
\draw (0,0)--(0,3);
\node[descmax] at (0,3) {$ 5 $};
\node[descvertex] at (0,2) {$ 3 $};
\node[descvertex] at (0,1) {$ 3 $};
\node[descleaf] at (0,0) {$ 2 $};
\draw[->,black!55] (.38,1.5)--(.85,1.5);
\begin{scope}[xshift=.58cm]
\draw[step=1,gray!22,very thin] (0,0) grid (3,3);
\draw[gray!65,dashed,thin] (0,0)--(3,3);
\draw[blue!70!black,very thick] (0,0)--(0,1)--(0,2)--(1,2)--(1,3)--(2,3)--(3,3);
\end{scope}\end{tikzpicture}\\[2mm]
$a=(0,1,1)$
\end{minipage}\hfill
\begin{minipage}[b]{.19\textwidth}\centering
\begin{tikzpicture}[x=.49cm,y=.49cm]
\draw (0,0)--(0,3);
\node[descmax] at (0,3) {$ 6 $};
\node[descvertex] at (0,2) {$ 4 $};
\node[descvertex] at (0,1) {$ 3 $};
\node[descleaf] at (0,0) {$ 2 $};
\draw[->,black!55] (.38,1.5)--(.85,1.5);
\begin{scope}[xshift=.58cm]
\draw[step=1,gray!22,very thin] (0,0) grid (3,3);
\draw[gray!65,dashed,thin] (0,0)--(3,3);
\draw[blue!70!black,very thick] (0,0)--(0,1)--(0,2)--(0,3)--(1,3)--(2,3)--(3,3);
\end{scope}\end{tikzpicture}\\[2mm]
$a=(0,1,2)$
\end{minipage}
\endgroup
\caption{The five three-edge chains in the class $\nu=2$ and their
Dyck paths from $(0,0)$ to $(3,3)$, using north and east steps
above the diagonal.  Read the labels from the leaf upward,
excluding the root, as $\ell_0,\ldots,\ell_{s-1}$ and set
$a_i=\ell_i-2$.  The nonmaximality conditions are precisely
$a_0=0$ and $0\le a_i\le a_{i-1}+1$; the $(i+1)$st north step
has $i-a_i$ preceding east steps.  This is an explicit bijection
for the endpoint class, not a claimed path bijection for the
intermediate statistic classes.}
\label{fig:gww-chain-dyck}
\end{figure}
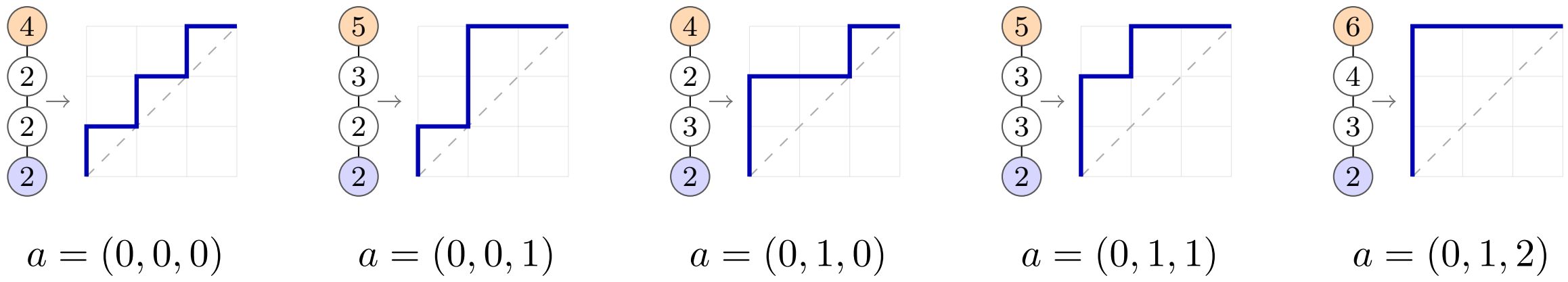

The numerator polynomials in the GWW recursion admit a concrete
conjectural interpretation by the description trees of type $(2,2)$
of Cori and Schaeffer~\cite{CoriSchaefferTrees2003}, counted by
OEIS A000309~\cite{OEISDescriptionTrees}.  Let $\mathcal T_s$ be
this family with $s\ge1$ edges: each leaf has label $2$, each
nonroot internal vertex $v$ has an integer label
$2\le\ell(v)\le2+\sum_{w\text{ child of }v}\ell(w)$, and the
root attains the upper bound.  Call an internal vertex maximal
when this upper bound is attained, and put
$\nu(\tau)=\#\{\text{leaves}\}+\#\{\text{maximal internal vertices}\}$;
the root therefore always contributes.  With the row indexing
$R_s(u)=u\sum_{n=1}^{s}T(s,n)u^{n-1}/(1+u)^{3s+1}$, the claim is
\begin{equation}\label{eq:gww-tree-refinement}
 B_s(z):=\sum_{n=1}^{s}(-1)^{n-1}T(s,n)z^{n-1}
 \ \stackrel{\mathrm{conj.}}{=}\
 \sum_{\tau\in\mathcal T_s}z^{\nu(\tau)-2},\qquad
 R_s(u)=\frac{u B_s(-u)}{(1+u)^{3s+1}}.
\end{equation}
Independent enumeration agrees with the differential recursion for
$1\le s\le9$; the first rows are $B_1=1$, $B_2=2+2z$,
$B_3=5+14z+5z^2$, and $B_4=14+74z+74z^2+14z^3$.
Thus the alternating signs of $T(s,n)$ would record a signed
statistic rather than negative cardinalities.  The conjecture
refines the row-sum relation
$B_s(1)=2^s(3s)!/[(s+1)!(2s+1)!]=\mathrm{A000309}(s)$ in this
indexing. The same unrefined numbers count rooted loopless
triangulations, for which Bernardi gave a bijective enumeration
using Kreweras walks~\cite{Bernardi2007}. This gives a walk connection
for the row sum; it does not identify the statistic $\nu$ on those
walks. Both extreme statistic classes have
$\mathrm{Cat}_s=(2s)!/[s!(s+1)!]$ members: $\nu=s+1$ assigns
the unique maximal labeling to every rooted plane tree shape,
whereas $\nu=2$ forces a chain whose nonmaximal labels encode a
Dyck path (Figures~\ref{fig:gww-description-s2} and
\ref{fig:gww-chain-dyck}); for the classical Catalan families of
plane trees and Dyck paths, see Ref.~\cite{Stanley2015Catalan}.
These endpoint counts follow directly
within the tree model; the refinement of all intermediate
columns remains conjectural.  Finally, if
$\mathcal B(q,z)=\sum_{s\ge1}q^sB_s(z)$ and $t=h^2/4$, the
GWW series~\eqref{eq:gseries} gives the exact formal identity
\begin{equation}\label{eq:gww-tree-GF}
 g(-z,2\sqrt t)+\frac z2
 =\frac{t}{1-z}+\frac{zt}{1-z}\,
 \mathcal B\!\left(\frac{t}{(1-z)^3},z\right).
\end{equation}
Under the conjecture, $q$ marks edges and $z$ marks $\nu-2$ in
$\mathcal B$.  This identifies the tree refinement inside the
algebraic GWW generating function while leaving a direct
statistic-preserving bijection with the matrix-integral
expansion as an additional problem.

\paragraph{An algebraic check of the finite tree comparison.}
For $q=2$, the uniform covering of
Ref.~\cite[Theorem C.1]{Voloshyn2026Hierarchies} provides a second
way to compute the GWW numerator polynomials. In the present
normalization, twice its reduced function, expanded in $h^2/4$,
reproduces $R_s$ through $s=9$. The positive numerator at $s=9$ has
coefficients
\[
 4862,\ 126008,\ 964264,\ 3027080,\ 4385332,\
 3027080,\ 964264,\ 126008,\ 4862 .
\]
These agree with the description-tree counts in
\eqref{eq:gww-tree-refinement}. The covering verifies the algebraic
side of the comparison independently of the differential recursion;
the general equality with the tree statistic remains conjectural.

\section{The algebraic curve and functional transformations of one-dimensional QCD}
\label{sec:qcdcurve}

\subsection{One-cut loop equation and the origin of the sextic}
\label{sec:qcd-spectral-origin}
The QCD curve can be derived directly from the planar loop equation.
For the exact determinant formulation and its 't Hooft--Veneziano
limit we refer to Ref.~\cite{BCVY2024QCD}, published in 2025. The
following calculation uses the same one-cut resolvent and
discriminant construction as \eqref{eq:gww-loop-equation}--\eqref{eq:gww-disc-origin},
with the exponential GWW weight replaced by the two fermion determinants.
Since $D_q^{\mathrm Q}=D_{-q}^{\mathrm Q}$, use the equivalent
insertion $(\det U)^{-q}$ and write $v=q/N=ku$.  Converting the
unitary Vandermonde to a holomorphic eigenvalue integral gives the
potential
\begin{equation}\label{eq:qcd-Penner-potential}
 V(\lambda)=(1+k+v)\log\lambda
 -k\log(1+h\lambda)-k\log(\lambda+h),\qquad
 V'=\frac{1+k+v}{\lambda}-\frac{k}{\lambda+h}
                         -\frac{k}{\lambda+h^{-1}}.
\end{equation}
The first logarithm includes the unitary measure.  Let
$\omega(\lambda)=\int\rho(\xi)/(\lambda-\xi)\,\dd\xi$
be normalized by $\omega\sim1/\lambda$ at infinity.  The planar
loop equation is $\omega^2-V'\omega+f=0$, where
$f=\int[V'(\lambda)-V'(\xi)]\rho(\xi)/(\lambda-\xi)\,\dd\xi$.
Put $a=h\omega(-h)$, $b=h^{-1}\omega(-h^{-1})$ and
$G=x\partial_x\Psi_{\mathrm Q}$.  Expansion at infinity and direct
differentiation of the determinant give
\begin{equation}
 a+b=u-1,\qquad G=\frac{k}{2}(1+b-a),\qquad
 a=\frac u2-\frac Gk,\quad b=\frac u2-1+\frac Gk.
\end{equation}
The vanishing $1/\lambda$ coefficient of $f$ eliminates
$\omega(0)$, leaving
\begin{equation}
 f(\lambda)=\frac{ka}{\lambda(\lambda+h)}
           +\frac{kb}{\lambda(\lambda+h^{-1})}.
\end{equation}
Consequently, with
\begin{align}
 D_h(\lambda)&=h\lambda^3+(1+h^2)\lambda^2+h\lambda,\\*
 N_h(\lambda)&=(1-k+v)h\lambda^2
 +(1+v)(1+h^2)\lambda+(1+k+v)h,
\end{align}
the squared spectral function is
\begin{align}\label{eq:qcd-spectral-quartic}
 (V'-2\omega)^2&=\frac{Q_h(\lambda)}{D_h(\lambda)^2},\\
 Q_h(\lambda)&=N_h(\lambda)^2-4D_h(\lambda)
 \left[(v-k)h\lambda+\frac v2(1+h^2)-kh^2+G(h^2-1)\right].\notag
\end{align}
On a regular one-cut branch the spectral numerator has the form
$(A\lambda+B)^2(\lambda-\lambda_1)(\lambda-\lambda_2)$.
Thus $\operatorname{Disc}_{\lambda}Q_h=0$.
For $E_0=k^2+(k+1)^2+v^2$, $z=(1-h^2)/(1+h^2)$ and
\begin{equation}\label{eq:qcd-W-from-G}
 W=\frac{zG-kz/2-k^2/2}{E_0},
\end{equation}
this discriminant is exactly
\begin{equation}\label{eq:qcd-disc-origin}
 \boxed{\quad
 \operatorname{Disc}_{\lambda}Q_h
 =2^{12}h^4E_0^6(1-h^2)^2(1+h^2)^6P_6(W).
 \quad}
\end{equation}
Here $P_6$ is the monic sextic below; its invariant arguments are
$r=[k^2(k+1)^2+(k^2+(k+1)^2)v^2]/E_0^2$,
$s=k^2(k+1)^2v^2/E_0^3$ and $t=z^2$.
The identity was checked as a full polynomial identity in $k,v,h,W$,
not through a truncated source series.  It supplies the missing
spectral derivation under the stated one-cut and planar-factorization
assumptions.  The discriminant condition includes other algebraic
sheets; the small-$h$ boundary condition and contour continuation
are still required to select a physical branch.  Colliding-pole
limits such as $h=1$ are taken by continuation, rather than by
cancelling vanishing factors at that point.

\subsection{The sextic in projective variables}

For the QCD triple in \eqref{eq:triples}, set
\begin{equation}\label{eq:qvariables}
 E_0=k^2+(k+1)^2+k^2u^2,\qquad
 z=\frac{1-h^2}{1+h^2},\qquad t=z^2,
\end{equation}
and form $r,s$ from the same symmetric expressions
\eqref{eq:rs}. The useful projective generating function is
\begin{equation}\label{eq:qW}
 W=\frac{kzR-k^2/2}{E_0}
   =\frac{k}{2E_0}(zhS_h-k).
\end{equation}
This definition contains both a multiplicative normalization and an
additive shift. Neither can be omitted when comparing the transformed
mass derivative with an invariant algebraic variable.

The resulting algebraic curve is the sextic
\begin{equation}\label{eq:P6}
 P_6(W)=W^3\left(W^3+\frac12W^2+\frac r4W+\frac s8\right)
 +\frac{tP_1(W)+t^2P_2(W)+t^3\Delta}{256}=0,
\end{equation}
where
\begin{align}\label{eq:P6parts}
P_1(W)={}&(16-192r)W^4-(64r+288s)W^3\notag\\
 &-(32r^2+120s)W^2-72rsW-27s^2,\\
P_2(W)={}&(-8r+48r^2-72s)W^2\notag\\
 &+(8r^2-48s+72rs)W+4r^3-18rs+54s^2.
\end{align}
The discriminant $\Delta$ in this formula is the symmetric invariant
in \eqref{eq:disc}. Unlike the oriented GWW quartic, the sextic does
not involve a choice of sign for $d$. This difference affects the
normalization of the invariant generating function, although the
projective parameter group is the same in both models.

For the positive fixed-sector germ, the small-$h$ boundary condition is
\begin{equation}\label{eq:qboundary}
 R(k,u,0)=\frac{u-1}{2},\qquad
 W(k,u,0)=\frac{ku-k-k^2}{2E_0}.
\end{equation}
Equivalently, in the chart $[1:M:u]$ with $M=1+1/k$,
\begin{equation}
 W(0)=\frac{(M-1)u-M}{2(1+M^2+u^2)}.
\end{equation}
The endpoint $h=0$ corresponds to $t=1$, not $t=0$. The distinction
matters because the polynomial degenerates differently on the two
mass slices:
\begin{align}
 P_6\big|_{t=0}&=W^3\left(W^3+\frac12W^2+\frac r4W+\frac s8\right),\\
 P_6\big|_{t=1}&=\frac{(4W+1)^2}{256}
 \left(16W^4-8rW^2-16sW+r^2-4s\right).
\end{align}
A repeated polynomial root on one of these slices does not specify
which regular branch approaches it. Boundary data and continuation
remain part of the fixed-sector definition.

For the expansions and limits below, set
\begin{equation}
 K=k(k+1),\qquad E=2K+1+v^2,
\end{equation}
and define a polynomial by clearing the projective denominators of the
sextic:
\begin{equation}\label{eq:rawQ}
 \mathcal Q(T;K,v,x)=16(1+x)^6E^6P_6(w;r,s,t),
\end{equation}
where
\begin{align}\label{eq:rawQsub}
 w&=\frac{2T+v(1-x)-K(1+x)}{2(1+x)E},&
 r&=\frac{K^2+(2K+1)v^2}{E^2},\\
 s&=\frac{K^2v^2}{E^3},&t&=\frac{(1-x)^2}{(1+x)^2}.
\end{align}
The relation to the original QCD derivative is
\begin{equation}
 T=(1-x)\left(\G_{\mathrm Q}-\frac v2\right)+kx.
\end{equation}
\subsection{Independent finite-\texorpdfstring{$N$}{N} Toeplitz check}
\label{sec:qcd-toeplitz-check}
For integer $N_f$ the exact determinant representation of one-dimensional
QCD~\cite{BCVY2024QCD} gives, for $0\le\ell\le N_f$,
\begin{equation}
 c_\ell(h)=\sum_{j=0}^{N_f-\ell}
 \binom{N_f}{j+\ell}\binom{N_f}{j}h^{2j+\ell},\qquad c_{-\ell}=c_\ell.
\end{equation}
Set $T_{ij}=c_{q+i-j}$ and
$A_{ij}=x\partial_x c_{q+i-j}$.  The independent observable is
\begin{equation}
 G_N=\frac1{N^2}\operatorname{Tr}(T^{-1}A),\qquad
 D_q^{\mathrm Q}=\det T.
\end{equation}
Calculations at 150 decimal digits give the following comparison.
The extrapolation fits $G_N=G_\infty+a/N^2+b/N^4$ using
$N=32,48,64$ in the first two rows and $N=24,36,48$ in the last.
\begin{center}\small
\begin{tabular}{ccccc}
\toprule
$(k,u,h)$&$N_{\max}$&$G_\infty$ from curve&Extrapolated $G_N$&Absolute difference\\\midrule
$(1,1/2,1/5)$&64&0.263458117377954&0.263458117377945&$9.3\,10^{-15}$\\
$(1,1/2,3/5)$&64&0.368638792493851&0.368638792488658&$5.2\,10^{-12}$\\
$(2,1/3,2/5)$&48&0.552902555246180&0.552902555236895&$9.3\,10^{-12}$\\\bottomrule
\end{tabular}
\end{center}
The finite matrices are constructed without using the sextic.
For example, at $(1,1/2,3/5)$ the unextrapolated $N=64$ value is
$0.368634327577591$.  These regular-point checks support the selected
physical branch in addition to the spectral derivation.  The quoted
differences are observed extrapolation differences, not rigorous
error bounds or a global saddle-dominance proof.

\subsection{Returning to the physical affine chart}

The physical QCD coordinates satisfy $b-a=1$. After a permutation or
sign change, let the raw coordinates be
$(\widetilde a,\widetilde b,\widetilde c)$. Restoring the physical
chart requires
\begin{equation}\label{eq:qrestore}
 \lambda=\frac1{\widetilde b-\widetilde a},\qquad
 k'=\lambda\widetilde a,\qquad k'u'=\lambda\widetilde c.
\end{equation}
Thus $E_0'=\lambda^2E_0$, whereas $r,s,t$, and the transported root
$W$ are unchanged. For pure transpositions the affine maps are
\begin{align}\label{eq:qmaps}
T_{ab}:&(k,u)\longmapsto\left(-k-1,\frac{ku}{k+1}\right),\\
T_{bc}:&(k,u)\longmapsto\left(\frac1{u-1},\frac{k+1}{k}\right),\\
T_{ac}:&(k,u)\longmapsto\left(\frac{ku}{k+1-ku},\frac1u\right).
\end{align}
The frequently used variant $u'=-ku/(k+1)$ in the first map
corresponds to the signed operation $(a,b,c)\mapsto(b,a,-c)$.
Both versions exchange $A_1$ and $A_2$. Their difference is a charge
sign change, which belongs to the kernel of the action on squared
coordinate labels. Keeping the signed lift explicit prevents a sign
convention from being mistaken for a different projective symmetry.

The defining relation \eqref{eq:qW} determines the transformed
mass derivative. Integrating that derivative at fixed external
parameters gives
\begin{equation}\label{eq:qStransform}
 S(k',u';h)=\frac{\lambda^2k}{k'}S(k,u;h)
 +\frac{k'^2-\lambda^2k^2}{k'}\ell(h)+C_T(k,u),\qquad
 \ell(h)=\log\frac{h}{1-h^2}.
\end{equation}
The logarithm follows from
$h\ell_h=(1+h^2)/(1-h^2)=z^{-1}$. The function $C_T$ is independent
of $h$ and carries the integration freedom left by the mass derivative.
Its value depends on the normalization and the continued logarithmic
sheets. The group relations impose corresponding consistency conditions
on these additive functions; the differential equation in $h$ alone
does not determine them.

Writing $T^*S=\alpha_TS+\beta_T\ell+C_T$, the coefficients are
\begin{equation}\label{eq:qalphabeta}
\begin{array}{c|c|c}
T&\alpha_T&\beta_T\\\hline
ab&-k/(k+1)&-(2k+1)/(k+1)\\
bc&1/[k(u-1)]&0\\
ac&1/[u(k+1-ku)]&k(u^2-1)/[u(k+1-ku)]
\end{array}
\end{equation}
The signed $ab$ variant has the same displayed coefficients when the
same algebraic root is transported. These affine relations concern the
analytic free energy reconstructed from $W$, not an independent
selection of a physical saddle at each transformed parameter value.

The sign operations can also be written directly in the physical chart.
For example, $b\mapsto-b$ gives
$k'=-k/(2k+1)$ and $u'=u$, while $c\mapsto-c$ gives $u'=-u$.
Together with \eqref{eq:qmaps}, these generate the full projective
signed-permutation group.

There is a further mass inversion. At finite $N$,
\begin{equation}
 D_q^{\mathrm Q}(h^{-1})=h^{-2NN_f}D_q^{\mathrm Q}(h),
\end{equation}
which follows by factoring $h^{-N}\det U$ and
$h^{-N}\det U^{-1}$ from the two determinants. In the normalization
\eqref{eq:qnorm}, this yields $S(h^{-1})=S(h)$ with consistent
logarithms. It sends $\G_{\mathrm Q}\mapsto k-\G_{\mathrm Q}$ and
$R\mapsto-R$, while $z\mapsto-z$; hence $t$ and $W$ are fixed.
This inversion commutes with the parameter permutations and has a
different origin from the affine chart transformations.

\subsection{A homogeneous potential for the free energy}

The affine law \eqref{eq:qStransform} suggests subtracting its
specific logarithmic term before discussing the free-energy operators.
Define
\begin{equation}\label{eq:qPhi}
 \Phi(k,u;h)=kS-k^2\ell(h),\qquad
 \widehat S=\frac{\Phi}{E_0}.
\end{equation}
Equation \eqref{eq:qStransform} becomes
$\Phi'=\lambda^2\Phi+k'C_T$. Thus $\widehat S$ is invariant up to
an $h$-independent function. This is exactly the ambiguity to which a
differential equation containing a mass derivative is insensitive.
Moreover,
\begin{equation}\label{eq:qPotentialDerivative}
 W=\frac{zh}{2}\partial_h\widehat S
   =t(t-1)\partial_t\widehat S.
\end{equation}
The logarithmic subtraction is therefore determined by the desired
projective derivative; it is not an arbitrary simplification of the
free energy.

To give $\Phi$ a homogeneous meaning away from the physical chart,
let $d_0=b-a$ and set
\begin{equation}\label{eq:qPhiLift}
 \Phi(a,b,c;h)=a(b-a)S\left(\frac a{b-a},\frac ca;h\right)-a^2\ell(h).
\end{equation}
This expression is homogeneous of degree two in $(a,b,c)$, or degree
one in $(A_1,A_2,A_3)$ on a chosen local square-root chart. The
homogeneity relation supplies the redundant scale equation when three
Euler derivatives are used to describe a system with only two
independent external parameter ratios.

\section{Differential recursions and the QCD coefficient pyramid}
\label{sec:qcddiff}

The QCD recurrence is most transparent after resumming the
small-charge series, rather than comparing successive powers of
the bare mass parameter. We first give the row and column forms of
the recurrence and describe the boundary data. The projective
derivation then relates the resulting PDE to the sextic.
Appendix~\ref{app:original-qcd-rec} retains the alternative
recursions for the original mass and free-energy coefficients.

\subsection{The small-charge chart and the row recursion}
\label{sec:pyramid-rows}

We now expand the selected root of the QCD sextic
\eqref{eq:P6} in the charge per color $v=ku$. The following
normalization organizes that expansion into a coefficient pyramid:
\begin{equation}\label{eq:pyr-vars}
 K=k(k+1),\qquad v=ku,\qquad x=h^2,\qquad
 \tau=\frac{Kx}{1-x},\qquad t_{\rm p}=4\tau,\qquad
 \zeta=-\frac{v}{1-x}.
\end{equation}
The resummed variable $t_{\rm p}$ is distinct from the invariant
mass coordinate $t$ in \eqref{eq:qvariables}. Define
\begin{align}\label{eq:pyr-definition}
 \G_{\mathrm Q}
 &=\frac v2+\frac{k^2x}{1-x}+(1-x)\mathcal A_{\rm p}(\zeta,t_{\rm p},x)
 =\frac v2+\frac{k^2x}{1-x}
   +\sum_{m\geq0}\frac{(-v)^{m+1}}{(1-x)^m}B_m(t_{\rm p},x),
 \notag\\
 \mathcal A_{\rm p}
 &=\sum_{m\geq0}\zeta^{m+1}B_m(t_{\rm p},x),
 \qquad B_m(t_{\rm p},x)=\sum_{j\geq0}A_{m,j}(t_{\rm p})x^j.
\end{align}
Here $\zeta,t_{\rm p},x$ are treated as independent in formal
expansions and differentiations. Expanding each $A_{m,j}$ in
$t_{\rm p}$ gives the third index of the array.

This GF is a root of the same sextic in a different chart. In the
cleared polynomial $\mathcal Q$ of \eqref{eq:rawQ}, substitute
\begin{equation}\label{eq:pyr-curve-sub}
 K=\frac{\tau(1-x)}x,\qquad v=-\zeta(1-x),\qquad
 T=(1-x)\tau+(1-x)^2a.
\end{equation}
The root is $a=\mathcal A_{\rm p}(\zeta,4\tau,x)$. Let
$\widetilde P(a;\zeta,t_{\rm p},x)$ be the polynomial obtained
after clearing denominators and removing parameter-only content,
with constant normalization fixed in \eqref{eq:pyr-special-quartic}
below. Thus the pyramid is an expansion of the QCD curve, with the
fixed-sector germ selecting its branch.

The main differential recurrence for this expansion is
\begin{equation}\label{eq:pyr-row-rec}
 \boxed{\mathscr L_{m+2}B_{m+2}=2E_m\mathscr M_mB_m,\qquad m\geq0,}
\end{equation}
with the two initial equations $\mathscr L_0B_0=\mathscr L_1B_1=0$.
Writing $\vartheta=t_{\rm p}\partial_{t_{\rm p}}$, the operators are
\begin{align}\label{eq:pyr-row-ops}
 E_m&=m-1+2\vartheta,\qquad H_m=m-\vartheta,\notag\\
 \mathscr L_m
 &=\bigl[t_{\rm p}(1-x)E_m-4x(H_m+1)\bigr]\partial_x
   +t_{\rm p}E_mH_m+2H_m(H_m+1),\notag\\
 \mathscr M_m&=2x(1-x)^2\partial_x+E_m+1+2xH_m-3mx^2.
\end{align}
Products denote composition; the rightmost operator acts first.
In particular, the equation differentiates both rows. Its
derivation is the coefficient extraction of the exact operator
identity \eqref{eq:pyr-cyclic-identity}, which follows from the
sextic PDE \eqref{eq:qInvariantPDE}. It is not a recurrence
inferred from the displayed low-order coefficients.

\subsection{The generating PDE and reconstruction by columns}
\label{sec:pyramid-columns}

Set $Q=\zeta\partial_\zeta$, $D_{\rm p}=Q+2\vartheta$ and
$J_b=Q-\vartheta-1$. On a term $\zeta^{m+1}B_m$,
$E_{Q-1}$ and $H_{Q-1}$ act as $E_m$ and $H_m$. Summing
\eqref{eq:pyr-row-rec}, including the two initial equations, gives
\begin{equation}\label{eq:pyr-generating}
 \boxed{\mathscr R_{\rm p}\mathcal A_{\rm p}=0,\qquad
 \mathscr R_{\rm p}=-\mathscr L_{Q-1}
         +2\zeta^2 E_{Q-1}\mathscr M_{Q-1}.}
\end{equation}
Conversely, the coefficient of $\zeta^{m+3}$ gives
\eqref{eq:pyr-row-rec}. In particular it is the factor $\zeta^2$,
placed to the left of the Euler operators, that shifts the row index by
two. The recurrence and the PDE therefore encode the same constraint.

For computation it is convenient to expand the same operator in $x$:
\begin{align}\label{eq:pyr-LM}
 \mathscr R_{\rm p}&=\sum_{i=0}^{3}x^iM_i\partial_x
                            +\sum_{i=0}^{2}x^iL_i,\notag\\
 M_0&=-t_{\rm p}(D_{\rm p}-2),\qquad
 M_1=(t_{\rm p}+4\zeta^2)(D_{\rm p}-2)+4(J_b+1),\notag\\
 M_2&=-8\zeta^2(D_{\rm p}-2),\qquad
 M_3=4\zeta^2(D_{\rm p}-2),\notag\\
 L_0&=2\zeta^2(D_{\rm p}-1)(D_{\rm p}-2)
                  -t_{\rm p}(D_{\rm p}-2)J_b-2J_b(J_b+1),\notag\\
 L_1&=4\zeta^2(D_{\rm p}-2)J_b,\qquad
 L_2=-6\zeta^2(Q-1)(D_{\rm p}-2).
\end{align}
These $L_i,M_i$ are column operators, distinct from the row operators
$\mathscr L_m,\mathscr M_m$. With
$\mathcal A_j=[x^j]\mathcal A_{\rm p}
=\sum_m\zeta^{m+1}A_{m,j}$, coefficient extraction now gives
\begin{equation}\label{eq:pyr-column-rec}
 \boxed{\begin{aligned}
 (j+1)t_{\rm p}(D_{\rm p}-2)\mathcal A_{j+1}
  ={}&(L_0+jM_1)\mathcal A_j\\
    &+[L_1+(j-1)M_2]\mathcal A_{j-1}\\
    &+[L_2+(j-2)M_3]\mathcal A_{j-2},\qquad j\geq0 .
 \end{aligned}}
\end{equation}
Columns of negative index vanish. In the physical formal class
$\zeta t_{\rm p}\mathbb Q[[\zeta,t_{\rm p}]]$ the Euler factor is
invertible term by term:
\begin{equation}\label{eq:pyr-inverse}
 (D_{\rm p}-2)^{-1}(\zeta^a t_{\rm p}^{b})
       =\frac{\zeta^a t_{\rm p}^{b}}{a+2b-2},\qquad a,b\geq1 .
\end{equation}
The right-hand side of \eqref{eq:pyr-column-rec} is first divided by
$t_{\rm p}$, then acted on by this inverse and divided by $j+1$.
Existence of the regular algebraic expansion supplies the required
divisibility for the chosen branch. No such divisibility is assumed
for arbitrary data in place of the boundary column. It follows that
the first column determines all other columns in this formal class.
The fixed-row version for computing individual entries is given in
Appendix~\ref{app:pyramid-checks}.

\subsection{The GWW and Narayana boundary triangles}
\label{sec:qcd-trees}

The recurrence \eqref{eq:pyr-column-rec} builds a three-index array.
The two boundaries discussed here are different specializations of
that array: the GWW column at $x=0$ and the Narayana triangle at
leading order in $t_{\rm p}$. The formal slice $x=1$ will give a
separate exact tree enumeration.

The first rows are
\begin{align}\label{eq:pyr-first}
 B_0&=\frac{1-\sqrt{1-t_{\rm p}}}{2},&
 B_1&=\frac{t_{\rm p}}{4(1-t_{\rm p})},\\
 B_2&=\frac{t_{\rm p}}{4(1-t_{\rm p})^{5/2}}
       +\frac{x t_{\rm p}}{4(1-t_{\rm p})^{3/2}},\notag\\
 B_3&=\frac{t_{\rm p}(1+t_{\rm p})}{4(1-t_{\rm p})^4}
       +\frac{3x t_{\rm p}}{4(1-t_{\rm p})^3}
       +\frac{x^2t_{\rm p}}{4(1-t_{\rm p})^2}.\notag
\end{align}
On the slice $k=1$, the first row also identifies a known integer
sequence. With the physical substitution $t_{\rm p}=8x/(1-x)$,
\[
 B_0\!\left(\frac{8x}{1-x},x\right)
   =\frac{1-\sqrt{(1-9x)/(1-x)}}{2}
   =2\bigl(G_{\rm M}(x)-1\bigr),\qquad
 G_{\rm M}(x)=\frac{5-\sqrt{(1-9x)/(1-x)}}4 .
\]
Muntingh obtained $G_{\rm M}$ as the generating function for the
number of terms in implicit divided differences
(\href{https://oeis.org/A162326}{OEIS A162326})~\cite[Prop.~1]{Muntingh2012}.
The identity concerns this row and this specialization.

The first column is precisely the GWW boundary, with elementary terms
removed. In the normalization of \eqref{eq:pyr-curve-sub}, the exact
specialization of the transformed sextic is
\begin{equation}\label{eq:pyr-special-quartic}
 \widetilde P(a;\zeta,t_{\rm p},0)
 =256t_{\rm p}^2
   P_4(a-\zeta/2+t_{\rm p}/4;-\zeta,\sqrt{t_{\rm p}}).
\end{equation}
Selecting the same analytic germ as in \eqref{eq:gseries} gives
\begin{equation}\label{eq:pyr-GWW}
 \mathcal A_{\rm p}(\zeta,t_{\rm p},0)
 =g(-\zeta,\sqrt{t_{\rm p}})+\frac\zeta2-\frac{t_{\rm p}}4.
\end{equation}
Inserting \eqref{eq:gseries} identifies this column with
$-\sum_{n\geq0}R_n(-\zeta)(t_{\rm p}/4)^{n+1}$, so the
GWW coefficient triangle, in the rational normalization of
Section~\ref{sec:gww}, supplies the data for
\eqref{eq:pyr-column-rec}. This is also the many-flavor limit of
the two matrix models, as explained in Section~\ref{sec:limit}.
Another exact boundary relation is
\begin{equation}\label{eq:pyr-second}
 A_{m,1}(t_{\rm p})=\frac{6t_{\rm p}}{m+1}
       \frac{\dd^2}{\dd t_{\rm p}^2}A_{m-2,0}(t_{\rm p}),
 \qquad m\geq2.
\end{equation}
In generating-function form this says
$Q\mathcal A_1=6t_{\rm p}\zeta^2\partial_{t_{\rm p}}^2\mathcal A_0$.
It follows by implicit differentiation of the transformed sextic
at $x=0$ and reduction modulo the boundary quartic
(Appendix~\ref{app:pyramid-checks}). The derivative suggests a pair
of marked components; a bijection would also have to account for
the prefactor and the charge-index shift.

For $m\geq1$ a useful normalization is
\begin{equation}\label{eq:pyr-N}
 B_m(t_{\rm p},x)=\frac{t_{\rm p}}{4(1-t_{\rm p})^{(3m-1)/2}}
          N_m\bigl(x(1-t_{\rm p});t_{\rm p}\bigr).
\end{equation}
Its boundary is the Narayana polynomial,
\begin{equation}\label{eq:pyr-Narayana}
 N_m(w;0)=\sum_{j=0}^{m-1}\frac1m
           \binom mj\binom m{j+1}w^j.
\end{equation}
In the classical lattice-path interpretation the coefficient of
$w^j$ counts Dyck paths of semilength $m$ with $j+1$
peaks~\cite[pp.~100--101]{Narayana1979}.
This boundary follows at all orders from the transformed sextic.
Write $\mathcal A_{\rm p}=t_{\rm p}\zeta Z/4+O(t_{\rm p}^2)$.
Inserting $a=t_{\rm p}\zeta Z/4$ in $\widetilde P$ gives
\begin{align}\label{eq:pyr-Nar-factor}
 [t_{\rm p}^2]\widetilde P(t_{\rm p}\zeta Z/4;\zeta,t_{\rm p},x)
 ={}&16\zeta^3x\,[\zeta xZ^2+\zeta(1-x)Z-Z+1]\notag\\
    &\times[\zeta^2(1-x)^2-2\zeta(1+x)+1]^2 .
\end{align}
Lower powers vanish in this substitution. The squared factor is
a formal unit near $\zeta=0$, and the regular branch obeys
$Z=1+\zeta Z(1-x+xZ)$. With $U=Z-1$ this is
$U=\zeta(1+U)(1+xU)$. Lagrange inversion yields, for $m\ge1$,
\begin{equation}\label{eq:pyr-Nar-invert}
 [\zeta^m]U=\frac{1}{m}[y^{m-1}](1+y)^m(1+xy)^m
           =\sum_{j=0}^{m-1}\frac1m\binom mj\binom m{j+1}x^j.
\end{equation}
Since $w=x$ at $t_{\rm p}=0$, this is \eqref{eq:pyr-Narayana}.
The leading resummed coefficient is meant here; setting $h=0$
at fixed $k$ in the original integral is a different substitution.

For instance the fourth normalized row is
\begin{equation}\label{eq:pyr-N4}
 N_4(w;4\tau)=1+13\tau+16\tau^2+(6+18\tau)w+(6+\tau)w^2+w^3 .
\end{equation}
At $\tau=0$ its coefficients are $(1,6,6,1)$, the fourth
Narayana row. The $\tau$ terms display how the interior of the
pyramid deforms this row.
\paragraph{Expansion about the Narayana boundary.}
Let $s=\sqrt{1-t_{\rm p}}$, $w=xs^2$, and $\xi=\zeta/s^3$. Set
\begin{equation}\label{eq:pyr-H-nar}
 H(\xi;w,t_{\rm p})=\sum_{m\ge1}N_m(w;t_{\rm p})\xi^m,\qquad
 \mathcal A_{\rm p}=\zeta\left[\frac{1-s}{2}+
                   \frac{t_{\rm p}s}{4}H\right].
\end{equation}
The substitution in the sextic of \eqref{eq:pyr-curve-sub} gives
\begin{equation}\label{eq:pyr-H-sextic}
 \mathcal P(H;\xi,w,t_{\rm p})=
 \frac{\widetilde P\bigl(\xi s^3[(1-s)/2+t_{\rm p}sH/4];
               \xi s^3,t_{\rm p},w/s^2\bigr)}
      {\xi^2t_{\rm p}^2s^{12}}=0.
\end{equation}
The quotient is a polynomial in $H,\xi,w,t_{\rm p}$: after cancellation
it contains only even powers of $s$. It is generically irreducible
of degree six in $H$. For example, the primitive specialization at
$(t_{\rm p},\xi,w)=(2,1,2)$ reduces, up to a unit, to the irreducible
polynomial $-H^6-H^5+2H^4-2H^3-2H+1$ over $\mathbb F_5$.
At $t_{\rm p}=0$ one has
\begin{align}
 \mathcal P(H;\xi,w,0)
 &=16\xi w\,\bigl(\xi(1+H)(1+wH)-H\bigr)
      \bigl(1-2\xi(1+w)+\xi^2(1-w)^2\bigr)^2 .\label{eq:pyr-H-nar-factor}
\end{align}
Thus the chosen branch begins with the Narayana series
$y=\xi(1+y)(1+wy)$. In what follows, set $z=wy$,
$A=(1+y)(1+z)$ and $D=1-yz$, so $\xi=y/A$.

The substitution $H=y+DV$, $\tau=D^3\lambda$ removes the
singular factors of the perturbation equation. In fact, put
\begin{equation}\label{eq:pyr-H-scaled}
 \mathcal R(V,\lambda;y,z)
  =\frac{A^6}{16D^6}
        \mathcal P(y+DV;y/A,z/y,4D^3\lambda).
\end{equation}
Direct expansion in \eqref{eq:pyr-H-sextic} gives
$\mathcal R\in\mathbb Z[y,z,V,\lambda]$ and
$\mathcal R(V,0;y,z)=zV(zV-1)$. For the divisibility check, write
$A^6\mathcal P(y+\eta;y/A,z/y,4\tau)/16$ as a polynomial in
$\tau,\eta$ over $\mathbb Z[y,z]$. The coefficient of $\tau^0\eta^j$
is divisible by $D^5$ for $j=1$ and by $D^4$ for $j=2$; the
other coefficients at $\tau^0$ vanish. For $\tau^1$ the required
powers of $D$ at $j=0,1,2,3$ are $3,2,1,0$, respectively, and
the remaining coefficients at $\tau^1$ vanish. All terms with
$\tau$-degree at least two already carry $D^6$ after scaling.

The coefficient recursion for $\mathcal R=0$ with $V(0)=0$ divides
only by $-z$. Consequently
\begin{equation}\label{eq:pyr-H-nar-series}
 H=y+\sum_{c\ge1}t_{\rm p}^{\,c}H_c(y,w),\qquad
 4^cH_c(y,w)=\frac{C_c(y,wy)}{(1-wy^2)^{3c-1}},
 \qquad C_c\in\mathbb Z[y,z].
\end{equation}
Indeed, initially $C_c=[\lambda^c]V$ belongs to
$\mathbb Z[y,z,z^{-1}]$. The original formal branch, expressed at
$\xi=y/[(1+y)(1+wy)]$, is regular in $w$. Since $D$ is a formal unit
at $w=0$, this rules out negative powers of $z$ in each $C_c$.
The factor $4^c$ is essential. For instance,
\begin{equation}\label{eq:pyr-H-first}
 4 H_1(y,w)
 =\frac{y^3\bigl(4+(1+6w+w^2)y+4w^2y^2\bigr)}{(1-wy^2)^2}.
\end{equation}

\paragraph{Degree bounds and finite-order positivity.}
At $\xi=0$ the polynomial in \eqref{eq:pyr-H-sextic} is
$Ht_{\rm p}(Ht_{\rm p}-4)$. Solving successively in $\xi$ yields
coefficients polynomial in $w$ and Laurent polynomial in $t_{\rm p}$;
regularity at $t_{\rm p}=0$ makes these polynomials in both variables.
To see the degree bounds, use a new series variable $\rho$ and a
small parameter $\epsilon$. The two rescalings
\[
  (\xi,w,H)=(\epsilon\rho,\epsilon^{-1},\epsilon U),
  \qquad
  (\xi,t_{\rm p},H)=(\epsilon\rho,\epsilon^{-1},
                         \epsilon\rho+\epsilon^2 U)
\]
give, respectively, the following lowest-order terms of
$\epsilon^{-1}\mathcal P$:
\begin{align*}
 &-4(\rho-1)^3\bigl((\rho-1)U+\rho\bigr)
             \bigl[t_{\rm p}(2\rho-1)^2+4\rho(1-\rho)\bigr],\\
 &4(\rho-1)^3\bigl[(1-\rho)(U-w\rho^2)-\rho^2\bigr].
\end{align*}
Both limiting equations have a simple branch at $\rho=0$ over the
corresponding parameter field. They give $U=\rho/(1-\rho)$ and
$U=w\rho^2+\rho^2/(1-\rho)$, respectively. Formal uniqueness
therefore gives $\deg_w N_m=m-1$, with leading coefficient one,
and $\deg_{t_{\rm p}}N_m=m-2$, with leading coefficient one,
for $m\ge3$. The small rows are $N_1=1$ and $N_2=1+w$.

For $1\le c\le6$, there are nonnegative integer polynomials $P_{c,j}$
such that
\begin{equation}\label{eq:pyr-H-positive}
 C_c(y,z)=\sum_{j=1}^{3c-1}P_{c,j}(y,z)D^{3c-1-j}.
\end{equation}
The certificate for $c=1$ is
$P_{1,2}=y^2(y^2+4yz^2+6yz+4y+z^2)$.
The ancillary file \nolinkurl{positivity_certificates_c1_c6.txt}
lists the explicit polynomials $P_{c,j}$ for $1\le c\le6$.
These decompositions make coefficientwise nonnegativity manifest:
each term $P_{c,j}(y,z)/(1-yz)^j$ has a nonnegative integer
power-series expansion in $y,z$.

The companion script \nolinkurl{verify_narayana_certificates.py}
reconstructs the scaled sextic and substitutes the right-hand sides
of \eqref{eq:pyr-H-positive} as coefficients of $\lambda^c$.
The residual vanishes through $\lambda^6$ in $\mathbb Z[y,z]$, so
formal uniqueness verifies all six identities. Substitution of
the Narayana series, which has nonnegative integer coefficients, proves
\begin{equation}\label{eq:pyr-H-positive-rows}
 [\tau^c]\,N_m(w;4\tau)\in\mathbb Z_{\ge0}[w],
       \qquad m\ge1,\quad 0\le c\le6 .
\end{equation}
The degree bound implies positivity of the full row for $m\le8$.
Direct exact extraction from \eqref{eq:pyr-H-sextic} verifies the
full rows for $m\le14$.

For the original rows $B_m$, a monomial $\tau^c w^j$ of
$N_m(w;4\tau)$ contributes a multiple of
\[
 \tau^{c+1}x^j (1-4\tau)^{-\frac{3m-1-2j}{2}},
 \qquad 0\le j\le m-1.
\]
The integer $3m-1-2j$ is positive, and
$(1-4\tau)^{-1/2}=\sum_{a\ge0}\binom{2a}{a}\tau^a$.
Thus $B_m$ has nonnegative integer coefficients through $\tau^7$
for every $m\ge1$, and to all orders in each fully positive row.
The coefficient of $\tau$ is the Narayana polynomial. The physical
substitution $\tau=Kx/(1-x)$ preserves these sign and integrality
statements in the corresponding powers of $K$. The exceptional row
$B_0=(1-\sqrt{1-4\tau})/2$ has nonnegative integer coefficients
by the Catalan formula.

\paragraph{The polar coefficient and the critical GWW column.}
Define the critical GWW series
\[
 G(X)=H(X;0,1)=\sum_{m\ge1}N_m(0;1)X^m .
\]
The specialization of \eqref{eq:pyr-H-sextic} is
\begin{equation}\label{eq:pyr-H-crit}
 J(U,X)=4XU^3+(1-24X)U^2+(30X-4)U+4X-27X^2,\qquad
 J(G(X),X)=0,\quad G(0)=0.
\end{equation}
There is an exact identity at the polar divisor:
\begin{equation}\label{eq:pyr-H-polar-identity}
 \mathcal R(V,\lambda;y,1/y)
   =\frac{1}{16}J\!\left(\frac{4V}{y},
                      \frac{4(1+y)^4}{y}\lambda\right).
\end{equation}
This follows by retaining terms of weight six in
\eqref{eq:pyr-H-scaled}, with $D,\tau,\eta$ of weights $1,3,1$.
It identifies the regular branch of the scaled equation with a
rescaled copy of the critical GWW series. In particular, for
every $c\ge1$,
\begin{equation}\label{eq:pyr-H-polar}
 4^cH_c
 =\frac{4^{c-1}y^{1-c}(1+y)^{4c}\,N_c(0;1)}
        {(1-wy^2)^{3c-1}}
      +O\bigl((1-wy^2)^{-(3c-2)}\bigr),
\end{equation}
where the Laurent expansion is taken at fixed generic $y$.

To evaluate the coefficient, parametrize \eqref{eq:pyr-H-crit} by
$X=(r^2-1)/(2r^3)$ and $G=2-(r+3/r)/2$ with $r=1$ at $X=0$.
Writing $R=(r^2-1)/2$ gives $R=X(1+2R)^{3/2}$ and
$\dd G/\dd R=(1-R)(1+2R)^{-3/2}$. Lagrange inversion then gives
\begin{equation}\label{eq:pyr-H-Kstar}
 N_m(0;1)=\frac1m[\rho^{m-1}](1-\rho)(1+2\rho)^{3(m-1)/2}
       =\frac{4^j\Gamma(3j+1)}{\Gamma(j+2)\Gamma(2j+2)},
       \qquad j=\frac{m-1}{2}.
\end{equation}
Thus \eqref{eq:pyr-H-nar-series} has pole order exactly $3c-1$
at generic points of the divisor. For $m=2j+1$ these coefficients
are the Kreweras excursion numbers (\href{https://oeis.org/A006335}{OEIS A006335}).
The entire odd GWW column has a related description: iterating
\eqref{eq:gRec} on monomials in \eqref{eq:pyr-GWW} gives
\[
 N_{2j+1}(0;t_{\rm p})
  =(1-t_{\rm p})^{3j+1}\sum_{n\ge0}
   \frac{(n+j+1)!(2n+2j+1)!}
        {(n+1)!(2n+1)!(j+1)!(2j+1)!}\,t_{\rm p}^n .
\]
By the numerator definition of
\href{https://oeis.org/A140136}{OEIS A140136}, this is its row
$j$. The identity \eqref{eq:pyr-H-polar-identity} relates the
deformation order to the index of the critical GWW column. It does
not require an exchange symmetry of the full coefficient array.

\subsection{An invariant second-order PDE}

The projective generating function obeys a compact equation in the
independent invariant coordinates $(r,s,t)$:
\begin{align}\label{eq:qInvariantPDE}
0=\mathscr D_{\mathrm{inv}}W:={}&2r(4r-1)W_{rr}
 +4s(6r-1)W_{rs}+18s^2W_{ss}\notag\\
 &+2t(t-1)W_{rt}+(4r-t)W_r+12sW_s.
\end{align}
The equation has total order two and contains no zeroth-order term.
Therefore the centered variable $W+1/12$ satisfies the same PDE. Its
validity for all regular roots of the sextic is established by the
polynomial-remainder calculation given in Appendix~\ref{app:verify}.
This is a full parameter identity rather than an inference from a
finite coefficient expansion.

Substituting $W=t(t-1)\widehat S_t$ gives a third-order equation for
the normalized potential itself:
\begin{align}\label{eq:qSinvariant}
0={}&2r(4r-1)\widehat S_{rrt}+4s(6r-1)\widehat S_{rst}
 +18s^2\widehat S_{sst}\notag\\
 &+2t(t-1)\widehat S_{rtt}
 +(4r+3t-2)\widehat S_{rt}+12s\widehat S_{st}.
\end{align}
The coefficient $4r+3t-2$ includes the derivative of the normalizing
factor $t(t-1)$. This term is needed when the equation is expressed
for the potential rather than for its derivative. In particular,
projective normalization of the free energy preserves the third order
of its equation; choosing its mass derivative as a new dependent
function is the separate operation that yields order two.

\subsection{The homogeneous operator triple}

Let $D_i=A_i\partial_{A_i}$ and
$\eta=t(t-1)$. The lift \eqref{eq:qPhiLift} has degree one in
$A_i$, so
\begin{equation}\label{eq:qEuler}
 (D_1+D_2+D_3)\Phi=\Phi,\qquad
 Y_{\mathrm Q}=E_0W=\eta\Phi_t.
\end{equation}
The parameter equation for $Y_{\mathrm Q}$ admits three operators of
identical form,
\begin{equation}\label{eq:qK}
 \mathscr K_i=D_i\left(D_i-1-\eta\partial_t+\frac t2\right),
 \qquad i=1,2,3.
\end{equation}
They commute: the Euler derivatives commute with each other and with
the common mass operator $-1-\eta\partial_t+t/2$. The complete
equation is the weighted cyclic relation
\begin{equation}\label{eq:qKcyclic}
 (A_2-A_3)\mathscr K_1Y_{\mathrm Q}
 +(A_3-A_1)\mathscr K_2Y_{\mathrm Q}
 +(A_1-A_2)\mathscr K_3Y_{\mathrm Q}=0.
\end{equation}
A coordinate permutation exchanges the three $\mathscr K_i$ and
the corresponding weights. The entire cyclic operator is alternating:
an odd permutation changes its sign while preserving its equation.

This decomposition can be checked before selecting an algebraic root.
On the chart $A_1=1$, $A_2=B$, $A_3=C$, a homogeneous degree-one
function is acted on by
\begin{equation}
 D_1=1-B\partial_B-C\partial_C,\qquad
 D_2=B\partial_B,\qquad D_3=C\partial_C.
\end{equation}
Let $E=1+B+C$ and $\delta_*= (1-B)(B-C)(C-1)$. With
$r=(B+C+BC)/E^2$ and $s=BC/E^3$, direct chain-rule substitution gives
\begin{equation}\label{eq:qOperatorIdentity}
 \cyc(A_j-A_k)\mathscr K_iY
 =-\frac{\delta_*}{2E}\,
 \mathscr D_{\mathrm{inv}}\left[\frac YE\right],
\end{equation}
where the invariant operator on the right is pulled back to the
$(B,C,t)$ chart. Equation \eqref{eq:qOperatorIdentity} holds for an
arbitrary function $Y$, with the invariant derivatives interpreted on
a regular local chart. It establishes that the cyclic triple is an
exact decomposition of the previously verified equation. It has not
been introduced by multiplying that equation by three unrelated
functions.

The corresponding operators for the free energy follow from
\begin{equation}
 \mathscr K_i[\eta\Phi_t]=\eta\mathscr T_i^{(t)}\Phi,
 \qquad
 \mathscr T_i^{(t)}=D_i\left(D_i-\eta\partial_t-\frac{3t}{2}\right)\partial_t.
\end{equation}
Each $\mathscr T_i^{(t)}$ has total order three. To return to
$x=h^2$, observe that
\begin{equation}
 \eta\partial_t=\chi(x)\partial_x,\qquad
 \chi(x)=\frac{x(1-x)}{1+x},\qquad Y_{\mathrm Q}=\chi(x)\Phi_x.
\end{equation}
Differentiating the mass-dependent prefactor gives, after a common
scalar normalization,
\begin{equation}\label{eq:qTx}
 \mathscr T_i^{(x)}=
 D_i[2(1+x)D_i+x-3-2x(1-x)\partial_x]\partial_x.
\end{equation}
More precisely,
$\mathscr K_i[\chi(x)\Phi_x]=\chi(x)\mathscr T_i^{(x)}\Phi/[2(1+x)]$.
The shifted free energy therefore satisfies
\begin{equation}\label{eq:qTcyclic}
 \cyc(A_j-A_k)\mathscr T_i^{(x)}\Phi=0.
\end{equation}
This is the desired permuted triple of third-order operators. The
literal permutation applies in homogeneous coordinates; restoring the
physical chart introduces conjugation by the degree-two scale factor,
consistent with the transformation of $\Phi$. Independent coordinate
sign changes leave the $A_i$ and their labels fixed, so the action of
the full $S_4$ on the triple factors through $S_3$.

\subsection{The projective action in pyramid coordinates}
\label{sec:pyramid-covariance}

The connection between the pyramid recurrence and the invariant
sextic PDE can be established at operator level. Let
$(a_1,a_2,a_3)=(a,b,c)$ and $d_0=b-a$. In homogeneous coordinates
put $E_h=a^2+b^2+c^2$, $z=(1-x)/(1+x)$, and
\begin{align}\label{eq:pyr-hom-lift}
 \mathcal H&=E_h W/z=f_0\mathcal A_{\rm p}+\ell_0,
 & f_0&=d_0^2(1-x),& \ell_0&=\tfrac12(cd_0-ab).
\end{align}
The physical chart is $d_0=1$; before restoring this chart the
pyramid arguments are $\zeta=-c/[d_0(1-x)]$ and
$t_{\rm p}=4abx/[d_0^2(1-x)]$. Conjugating \eqref{eq:qK} by $z$
gives
\begin{equation}\label{eq:pyr-hom-ops}
 \overline{\mathscr J}_i
  = D_i(D_i-\tfrac12-\chi\partial_x),\qquad
 \mathscr J_i=f_0^{-1}\overline{\mathscr J}_i f_0 .
\end{equation}
The second equation is operator conjugation. The mass derivative in
the homogeneous expression holds the three coordinates fixed.
Each $\overline{\mathscr J}_i$ annihilates $ab,ac,bc$ separately:
on these monomials the squared-coordinate Euler derivative $D_i$
has eigenvalue zero or $1/2$. In particular the affine term
$\ell_0$ does not introduce a source into the equation for
$\mathcal A_{\rm p}$.

For explicit comparison with \eqref{eq:pyr-generating}, put
$\delta=2k+1=\sqrt{1+t_{\rm p}(1-x)/x}$ on the physical chart,
retaining its sheet under parameter transformations. The operators
in pyramid variables are
\begin{align}\label{eq:pyr-dX}
 \mathfrak d_1&=\frac{\delta}{4}(D_{\rm p}-2)-\frac{Q-2}{4},
 &\mathfrak d_2&=-\frac{\delta}{4}(D_{\rm p}-2)-\frac{Q-2}{4},
 &\mathfrak d_3&=\frac Q2,\notag\\
 X&=\chi\partial_x+\frac{\vartheta+xQ}{1+x},
 & \mathscr J_i&=\mathfrak d_i
           \left(\mathfrak d_i-\frac12-X+\frac{x}{1+x}\right).
\end{align}
Here $X$ represents $\chi\partial_x$ at fixed $(k,v)$, whereas
the displayed $\partial_x$ holds $(\zeta,t_{\rm p})$ fixed.
The terms $\vartheta+xQ$ are necessary for the chain rule.
With weights $w_i$ as in \eqref{eq:qweights}, expansion of
\eqref{eq:pyr-dX} gives the exact identity
\begin{equation}\label{eq:pyr-cyclic-identity}
 \boxed{\mathscr C_{\rm p}:=\sum_i w_i\mathscr J_i
       =\frac{\delta(1-x)}{8(1+x)}\mathscr R_{\rm p}.}
\end{equation}
The weights multiply externally. Both sides have the same action
on an arbitrary test function, before choosing a root. Combining
\eqref{eq:qOperatorIdentity}, \eqref{eq:pyr-hom-lift} and
\eqref{eq:pyr-cyclic-identity} derives \eqref{eq:pyr-row-rec}
from the sextic. The row projection and exact verification are
given in Appendix~\ref{app:pyramid-checks}.

The homogeneous operators in \eqref{eq:pyr-hom-ops} are permuted
by $S_4$, with the label action through $S_4/V_4\simeq S_3$.
For a signed permutation $P$, let $(a',b',c')=P(a,b,c)$,
$d_0'=b'-a'$ and $\rho=(d_0/d_0')^2$. On the transported
sheet, the affine-chart laws are
\begin{align}\label{eq:pyr-cov}
 P^*\mathscr J_i(P^*)^{-1}
   &=\rho\,\mathscr J_{\sigma(i)}\rho^{-1},\notag\\
 P^*\mathcal A_{\rm p}
   &=\rho\mathcal A_{\rm p}
    +\frac{cd_0-ab-c'd_0'+a'b'}{2(d_0')^2(1-x)}.
\end{align}
The affine part divided by $\rho$ lies in the common kernel
of the $\mathscr J_i$, by the bilinear-monomial observation above.
Thus the operator actions transform covariantly even though the
regular pyramid GF is not an invariant scalar. This is an action on
the generating system and on its charts, not an asserted permutation
of the individual rows of one physical series.

A useful alternative takes divided differences. On the physical
chart put $b_i=a_i^2$ and, for cyclic $(i,j,k)$, set
\begin{equation}\label{eq:pyr-Z}
 \mathscr Z_i=\frac{\mathscr J_j-\mathscr J_k}{b_j-b_k}.
\end{equation}
The denominator acts by left multiplication. Away from coinciding
squared coordinates, the equality of the three actions on
$\mathcal A_{\rm p}$ is equivalent to the single weighted equation
$\mathscr C_{\rm p}\mathcal A_{\rm p}=0$. In fact,
\begin{equation}\label{eq:pyr-Zdiff}
 \mathscr Z_1-\mathscr Z_3=-\frac{\mathscr C_{\rm p}}{w_1w_3},
 \qquad
 \mathscr Z_2-\mathscr Z_3=\frac{\mathscr C_{\rm p}}{w_2w_3}.
\end{equation}
This reformulation does not imply pairwise equality of the original
components $\mathscr J_i$. Its relation to the standard GWW
third-order operators is established in Section~\ref{sec:operator-limit}.

\subsection{Exact tree enumeration on the collapsed slice}
A complete tree interpretation is available after summing over the
explicit $x$ statistic. Define
\begin{equation}\label{eq:pyr-collapsed}
 \mathcal T_{\mathrm Q}(\zeta,\tau)
 =\mathcal A_{\rm p}(\zeta,4\tau,1)
 =\sum_{m,a\geq0}q(m,a)\zeta^{m+1}\tau^{a+1}.
\end{equation}
In this formula $x=1$ is a formal specialization of the stripped
rows $B_m$, with $\tau$ held independent. It is not the physical
massless limit $h=1$ at fixed $k$, for which \eqref{eq:pyr-vars}
is singular. The exact coefficient formula is
\begin{equation}\label{eq:pyr-product}
 \boxed{\displaystyle
 q(m,a)=\frac{4^{m+a}
       \bigl((a+1)/2\bigr)_m\bigl((m+1)/2\bigr)_a}
       {(m+1)!(a+1)!},\qquad m,a\geq0.}
\end{equation}
Here $(b)_n=b(b+1)\cdots(b+n-1)$ and $(b)_0=1$.
The first entries are
\begin{equation}\label{eq:pyr-table}
 \bigl(q(m,a)\bigr)_{0\leq m,a\leq3}
 =\begin{pmatrix}
 1&1&2&5\\
 1&4&16&64\\
 2&16&100&560\\
 5&64&560&4096
 \end{pmatrix}.
\end{equation}
In particular $q(m,0)=\mathrm{Cat}_m$ and $q(m,1)=4^m$.
The symmetry $q(m,a)=q(a,m)$ is manifest in
\eqref{eq:pyr-product}; the following construction explains both
that symmetry and the integrality of the formula.

Consider unrooted bipartite plane trees with at least one edge. Each
vertex of degree $d$ carries a distinguished corner and a full plane
binary tree with $d$ leaves. The cyclically ordered incident edges,
read from the distinguished corner, correspond in order to the binary
tree's leaves. There are $\mathrm{Cat}_{d-1}$ binary decorations;
a leaf of the original tree has a unique corner and decoration.
Isomorphisms preserve the colors, the plane embedding, the corners,
and the decorations. Then $q(m,a)$ counts these objects with
$m+1$ black vertices and $a+1$ white vertices. No global root is
chosen. The corner at each vertex is a local mark, not an additional
root of the whole tree.

\begin{figure}[htbp]
\centering
\begin{tikzpicture}[x=1cm,y=1cm,line width=.8pt]
 \begin{scope}[shift={(-3.2,1.1)}]
  \draw (-1.35,0)--(1.35,0);
  \foreach \xx in {-1.35,.45} \fill (\xx,0) circle (2.7pt);
  \foreach \xx in {-.45,1.35} \draw[fill=white] (\xx,0) circle (2.7pt);
  \fill[orange!85!black] (-.45,.16)--(-.58,.42)--(-.32,.42)--cycle;
  \fill[orange!85!black] (.45,.16)--(.32,.42)--(.58,.42)--cycle;
  \node[font=\small] at (0,-.47) {$(+,+)$};
 \end{scope}
 \begin{scope}[shift={(3.2,1.1)}]
  \draw (-1.35,0)--(1.35,0);
  \foreach \xx in {-1.35,.45} \fill (\xx,0) circle (2.7pt);
  \foreach \xx in {-.45,1.35} \draw[fill=white] (\xx,0) circle (2.7pt);
  \fill[orange!85!black] (-.45,.16)--(-.58,.42)--(-.32,.42)--cycle;
  \fill[orange!85!black] (.45,-.16)--(.32,-.42)--(.58,-.42)--cycle;
  \node[font=\small] at (0,-.72) {$(+,-)$};
 \end{scope}
 \begin{scope}[shift={(-3.2,-1.0)}]
  \draw (-1.35,0)--(1.35,0);
  \foreach \xx in {-1.35,.45} \fill (\xx,0) circle (2.7pt);
  \foreach \xx in {-.45,1.35} \draw[fill=white] (\xx,0) circle (2.7pt);
  \fill[orange!85!black] (-.45,-.16)--(-.58,-.42)--(-.32,-.42)--cycle;
  \fill[orange!85!black] (.45,.16)--(.32,.42)--(.58,.42)--cycle;
  \node[font=\small] at (0,-.72) {$(-,+)$};
 \end{scope}
 \begin{scope}[shift={(3.2,-1.0)}]
  \draw (-1.35,0)--(1.35,0);
  \foreach \xx in {-1.35,.45} \fill (\xx,0) circle (2.7pt);
  \foreach \xx in {-.45,1.35} \draw[fill=white] (\xx,0) circle (2.7pt);
  \fill[orange!85!black] (-.45,-.16)--(-.58,-.42)--(-.32,-.42)--cycle;
  \fill[orange!85!black] (.45,-.16)--(.32,-.42)--(.58,-.42)--cycle;
  \node[font=\small] at (0,-.72) {$(-,-)$};
 \end{scope}
\end{tikzpicture}
\caption{The four decorated bipartite plane trees counted by
$q(1,1)=4$: two black vertices and two white vertices. The underlying
tree must be a path. Each of its two degree-two vertices has a unique
binary decoration and two choices of distinguished corner, indicated
above or below the path. Leaf corners and their trivial decorations
are suppressed. Plane isomorphisms preserve cyclic order; reflections
are not identified. The resulting contribution is
$4\zeta^2\tau^2$.}
\label{fig:qcd-four-trees}
\end{figure}
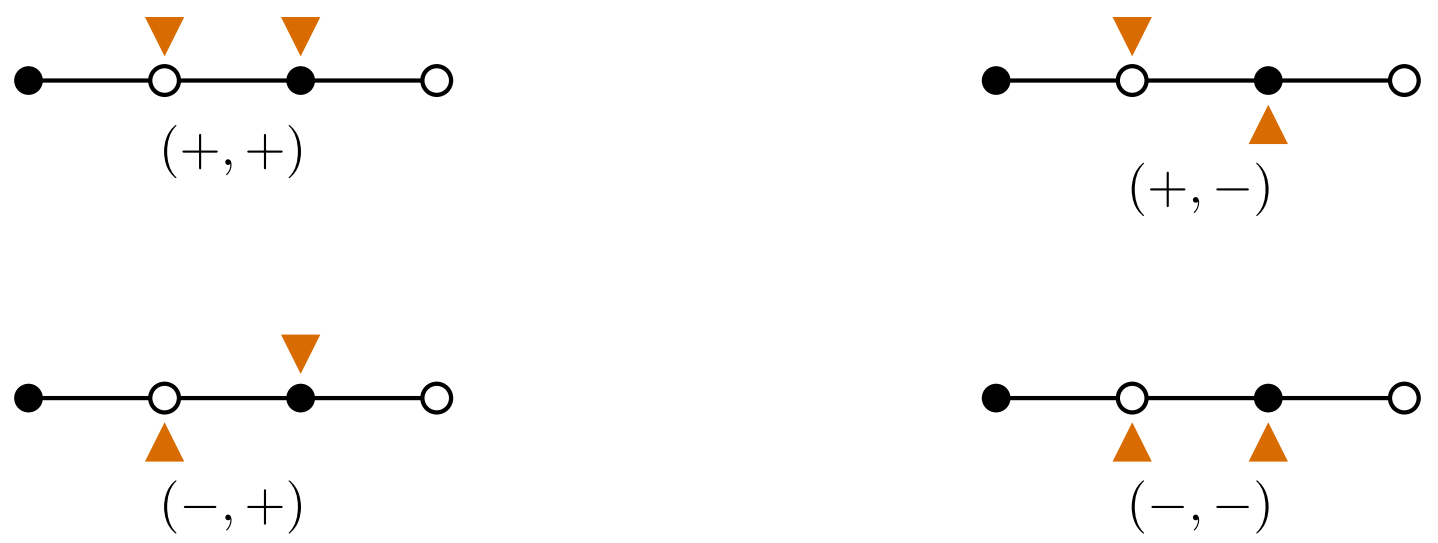

For a planted tree the parent edge fixes a reference in the cyclic
order. A vertex with $n$ children has degree $n+1$ and therefore
$(n+1)\mathrm{Cat}_n=\binom{2n}{n}$ local choices. Consequently,
the two planted generating functions satisfy
\begin{equation}\label{eq:pyr-planted}
 U=\frac{\zeta}{\sqrt{1-4V}},\qquad
 V=\frac{\tau}{\sqrt{1-4U}},\qquad U(0,0)=V(0,0)=0.
\end{equation}
The weights $\zeta$ and $\tau$ mark black and white vertices,
respectively. Removing the global root by the dissymmetry identity
for enriched trees~\cite{LabelleLeroux1996} gives
\begin{equation}\label{eq:pyr-tree-GF}
 \boxed{\displaystyle
 \mathcal T_{\mathrm Q}
 =\frac\zeta2\bigl(1-\sqrt{1-4V}\bigr)
  +\frac\tau2\bigl(1-\sqrt{1-4U}\bigr)-UV.}
\end{equation}
The two positive terms count vertex-rooted trees, separated by root
color; $UV$ counts trees with a distinguished edge. Color exchange
interchanges $U,V$ and $\zeta,\tau$, giving the coefficient symmetry
directly. Figure~\ref{fig:qcd-local-decoration} shows the local
decoration, and Figure~\ref{fig:qcd-four-trees} gives the complete
example $q(1,1)=4$. The algebraic identification with the QCD curve
and the coefficient extraction are proved in
Appendix~\ref{app:qcd-trees}.

\begin{figure}[htbp]
\centering
\begin{tikzpicture}[x=1cm,y=1cm,line width=.65pt]
 \draw (-4.3,0)--(-4.3,1.2);
 \draw (-4.3,0)--(-5.34,-.6);
 \draw (-4.3,0)--(-3.26,-.6);
 \fill[orange!85!black] (-4.3,0)--(-4.19,.28)--(-4.00,-.02)--cycle;
 \fill[black] (-4.3,0) circle (2.5pt);
 \node[font=\small] at (-4.3,1.55) {degree $3$};
 \node[font=\small,align=center] at (-4.3,-1.1)
   {marked corner selects\\an order of the three edges};
 \node at (-1.9,.35) {$\longleftrightarrow$};
 \draw (-.5,.9)--(-1.05,.2)--(-1.40,-.5);
 \draw (-1.05,.2)--(-.70,-.5);
 \draw (-.5,.9)--(.20,-.5);
 \foreach \p in {(-1.40,-.5),(-.70,-.5),(.20,-.5)}
   \fill \p circle (1.5pt);
 \node[font=\scriptsize] at (-1.4,-.77) {$1$};
 \node[font=\scriptsize] at (-.7,-.77) {$2$};
 \node[font=\scriptsize] at (.2,-.77) {$3$};
 \node at (1.35,.3) {or};
 \draw (3.10,.9)--(2.40,-.5);
 \draw (3.10,.9)--(3.65,.2)--(3.30,-.5);
 \draw (3.65,.2)--(4.00,-.5);
 \foreach \p in {(2.40,-.5),(3.30,-.5),(4.00,-.5)}
   \fill \p circle (1.5pt);
 \node[font=\scriptsize] at (2.4,-.77) {$1$};
 \node[font=\scriptsize] at (3.3,-.77) {$2$};
 \node[font=\scriptsize] at (4,-.77) {$3$};
 \node[font=\small] at (1.3,1.55) {two binary decorations: $\mathrm{Cat}_2=2$};
\end{tikzpicture}
\caption{A vertex of degree three has two full plane binary-tree
decorations. The orange mark distinguishes a corner and orders the
incident edges, which are matched to the binary leaves. When a parent
edge is fixed, there are three possible positions for this mark and
two decorations, hence $3\mathrm{Cat}_2=6=\binom42$ local choices.
The binary trees are decorations of one vertex, not extra vertices
counted by $\zeta$ or $\tau$.}
\label{fig:qcd-local-decoration}
\end{figure}
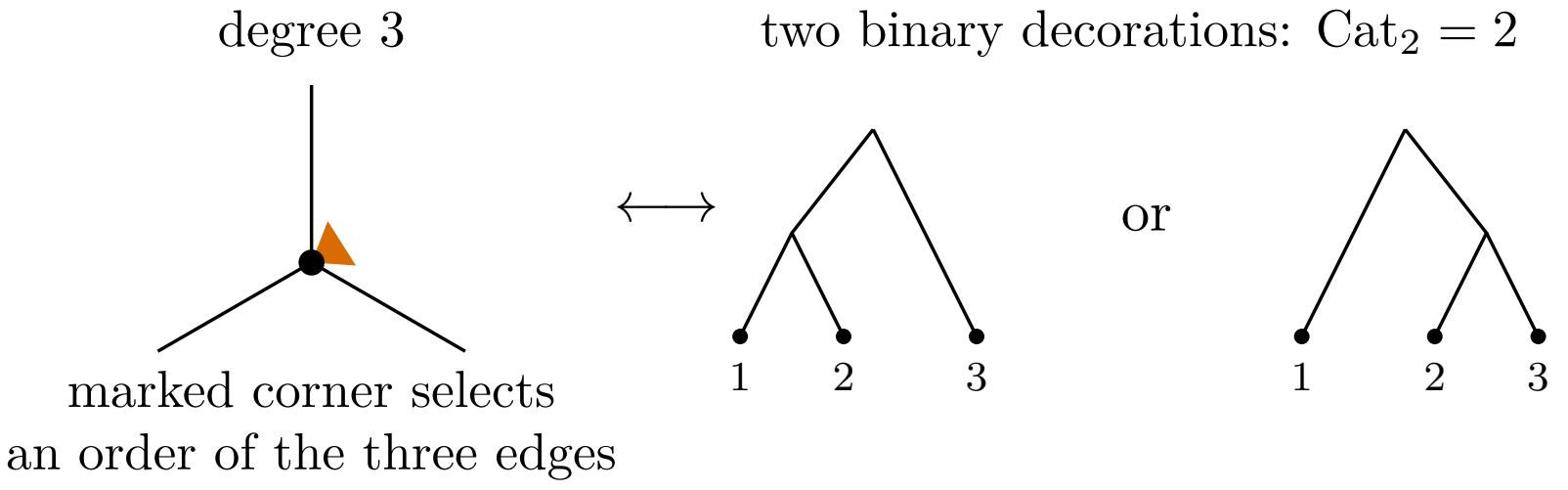

This result counts a formal coefficient slice of the analytic QCD
generating function. It does not identify these trees with the original
gauge configurations, nor does it imply that the full refined pyramid
has already been enumerated. The GWW description-tree refinement in
\eqref{eq:gww-tree-refinement} and the present decorated bipartite
trees therefore have different logical status: the former statistic
identity remains conjectural here, whereas \eqref{eq:pyr-product}
and \eqref{eq:pyr-tree-GF} are exact for the collapsed QCD slice.

\section{The many-flavor limit and the relation between the two models}
\label{sec:limit}

We connect the quartic of Section~\ref{sec:gww} to the QCD
sextic of Section~\ref{sec:qcdcurve} at the level of the matrix
integral, the algebraic equation, and the pyramid recurrence.
The coefficient limit and the operator limit must be taken
with the same rescaled charge and source.

\subsection{The matrix-integral limit and the free-energy normalization}

The relation between the models is obtained by retaining the charge per
color rather than the charge per flavor. Let
\begin{equation}\label{eq:limit}
 k\longrightarrow\infty,\qquad
 u_{\mathrm Q}=\frac vk,\qquad h_{\mathrm Q}=\frac H{2k},
\end{equation}
with $v,H$ fixed. At the level of the potential,
\begin{equation}
 k\log(1+h_{\mathrm Q}U)=\frac H2U+O(k^{-1}),
\end{equation}
and the same expansion holds for $U^{-1}$. The two exact determinant
factors in the QCD integral therefore tend to the GWW exponential. The
sector insertion remains fixed because $q/N=ku_{\mathrm Q}=v$.
With the static mass factor removed and the corresponding branch
continued consistently, this gives
\begin{equation}\label{eq:freeLimit}
 \Psi_{\mathrm Q}\left(k,\frac vk,\frac H{2k}\right)
 \longrightarrow\Psi_{\mathrm G}(v,H),\qquad
 \G_{\mathrm Q}\longrightarrow\G_{\mathrm G}.
\end{equation}
The normalization in this statement matters. Since
$\Psi_{\mathrm Q}=k(S+\log h_{\mathrm Q})$, the free energy $S$
itself does not tend directly to $\Psi_{\mathrm G}$. Likewise,
$R\to-1/2$, and the finite information is carried by
$k(R+1/2)=\G_{\mathrm Q}$. An unrescaled comparison of $R$ would
lose the GWW generating function.

The determinant expansion gives the matrix-integral origin of the
limit. When it is used for a planar branch, one must retain the branch
that continues the relevant finite-source germ. A theorem about
uniform convergence across a change of dominant saddle would require
additional control. The algebraic limit below identifies the curves
and their local branches without imposing such a global assertion.

\subsection{The degeneration of the projective triples}

The naive projective limit of the QCD triple is
\begin{equation}
 [k:k+1:ku_{\mathrm Q}]=[k:k+1:v]\longrightarrow[1:1:0],
 \qquad t\longrightarrow1.
\end{equation}
Thus simply taking a limit of $r,s$ erases the finite charge $v$ and
the source $H$. The GWW variables are transverse data near this
degenerate point. They are retained by the rescaled triple
\begin{equation}\label{eq:transverse}
 [i(b-a):c:2ah_{\mathrm Q}]=[i:v:H],
\end{equation}
which becomes $[v:H:i]$ after a permutation.

Equation \eqref{eq:transverse} explains both the similarity and the
difference of the two projective descriptions. They are related through
a degeneration with a retained transverse scale, rather than by a
regular identification of two generic parameter triples. In geometric
terms this is the local rescaling associated with a blow-up of the
degenerate point. In particular, a GWW permutation exchanging source and
charge cannot be identified with an ordinary permutation of the generic
QCD triple while leaving its additional mass coordinate untouched.
The source enters the transverse data of the limit.

\subsection{An explicit degeneration of the sextic to the quartic}

Use the cleared sextic \eqref{eq:rawQ}. The reduction in algebraic
degree can be verified directly, without radical expressions.
For a formal generating-function variable $g$, substitution of
$x=H^2/(4k^2)$ gives the exact coefficientwise limit
\begin{equation}\label{eq:polyLimit}
 \lim_{k\to\infty}k^{-4}
 \mathcal Q\left((1-x)(g-v/2)+kx;k(k+1),v,x\right)
 =16P_4(g;v,H).
\end{equation}
The right-hand polynomial is exactly \eqref{eq:P4} with $u=v$ and
$h=H$. This identity was obtained by a full expansion of the cleared
polynomial; it does not depend on a finite truncation of a root series.
The leading polynomial in the scaled generating-function variable has
degree four. The other two generic sextic roots do not remain finite
roots in this scaling, so the degree reduction is consistent with the
six-sheeted curve before degeneration.

\subsection{The coefficient limit and the order of the GWW recursion}

The auxiliary QCD coefficients recover the GWW sequence through
\begin{equation}\label{eq:fLimit}
 R_n(v)=\lim_{k\to\infty}\frac{f_n(k,v/k)}{k^{2n}},
 \qquad n\geq0.
\end{equation}
For the first two terms, the limit gives
$R_0=v/(1+v)$ and $R_1=v/(1+v)^4$ immediately from
\eqref{eq:qfirstf}. Together with the source rescaling in
\eqref{eq:limit}, the coefficient relation
\eqref{eq:qCfromf} also reproduces the regular GWW free-energy
series. At order $H^{2n}$, the leading contribution for $n\geq2$
comes from $f_{n-1}$, while the lower auxiliary coefficients have lower
powers of $k$. The first $H^2$ coefficient must be retained separately,
since its elementary contribution combines with $f_0$.

The differential-order comparison is more subtle than the coefficient
limit alone suggests. In the variables $K,v$, write
\begin{equation}
 f_n(K,v)=K^nR_n(v)+K^{n-1}S_n(v)+\cdots.
\end{equation}
Coefficients multiplying parameter derivatives can raise the power of
$K$. Consequently, the leading equation obtained from the QCD
second-order constraint can involve both $R_n$ and $S_n$. The first
nontrivial example is
\begin{equation}
 f_1(K,v)=K\frac{v}{(1+v)^4}-\frac{v^2}{(1+v)^3},
\end{equation}
so that $S_1=-v^2/(1+v)^3$. The leading parameter relation contains
\begin{align}\label{eq:mixing}
0={}&v^2(1-v^2)R_1''-2v(3v^2+2)R_1'
 +6(1-v^2)R_1\notag\\
&+2vS_1'-2vR_0'.
\end{align}
The $S_1$ contribution does not vanish. Dropping it does not yield a
closed second-order raising equation for $R_1$. This example
explains why the QCD second-order parameter constraint is compatible
with the GWW third-derivative recursion. The resummed chart below
makes the transverse contribution explicit at generating-function level.

\subsection{How the generating systems meet at the GWW boundary}
\label{sec:operator-limit}

The many-flavor scaling has two complementary effects. At the
algebraic level, \eqref{eq:polyLimit} reduces the QCD sextic
to the GWW quartic. In the small-charge chart the same scaling is
$x\to0$ with $\zeta,t_{\rm p}$ fixed and $H=\sqrt{t_{\rm p}}$.
The limit of the stripped QCD derivative is therefore
$\mathcal A_0$, the GWW column in \eqref{eq:pyr-GWW}. Conversely,
once this boundary column is specified, \eqref{eq:pyr-column-rec}
generates the regular transverse coefficients of the QCD pyramid.
This gives a coefficient-level connection, in addition to the
limit of the matrix potentials. In particular, the first transverse
column is given exactly by \eqref{eq:pyr-second}.

The operator limit retains this distinction between boundary and
transverse information. From \eqref{eq:pyr-dX} and \eqref{eq:pyr-Z},
\begin{equation}\label{eq:pyr-Z3}
 \mathscr Z_3=\tfrac14(D_{\rm p}-2)
       \left(Q-1+2X-\frac{2x}{1+x}\right).
\end{equation}
At $x\to0$, $\delta\to\infty$, $X\to\vartheta$ on regular jets,
and $w_1\sim\delta^2/4$, $w_2\sim-\delta^2/4$, $w_3=-\delta$.
Equations \eqref{eq:pyr-cyclic-identity} and \eqref{eq:pyr-Zdiff}
therefore show that all three $\mathscr Z_i$ approach
\begin{equation}\label{eq:pyr-Zlim}
 \mathscr Z_0=\tfrac14(D_{\rm p}-2)(D_{\rm p}-1).
\end{equation}
For the undivided normalized actions
$\mathcal U_i=\mathscr J_i\mathcal A_{\rm p}/E_0$, the limit is
$(\mathscr Z_0\mathcal A_0/2,\mathscr Z_0\mathcal A_0/2,0)$:
$\mathscr J_3$ remains finite and $b_1/E_0,b_2/E_0\to1/2$.
The two differences multiplied by $\delta^2$ approach the same
boundary operator, $\mathscr R_{\rm p}|_{x=0}/2$, which acts as
$(L_0\mathcal A_0-t_{\rm p}(D_{\rm p}-2)\mathcal A_1)/2$.
It is not an equation for $\mathcal A_0$ alone. Together with
$Q\mathcal A_1=6t_{\rm p}\zeta^2\partial_{t_{\rm p}}^2\mathcal A_0$
it gives the closed third-order identity
\begin{equation}\label{eq:pyr-boundary-PDE}
 \left[ Q L_0-6t_{\rm p}(D_{\rm p}-2)
           \circ\bigl(t_{\rm p}\zeta^2\partial_{t_{\rm p}}^2\bigr)
 \right]\mathcal A_0=0 .
\end{equation}
The extra differential order comes from eliminating the first
transverse coefficient, not from a regular limit of second-order
operators.

For $v=-\zeta$, the centered GWW variable is
$\tau_{\rm G}=-4\mathcal A_0+2\zeta-(1+\zeta^2+t_{\rm p})/2$,
in the normalization of \eqref{eq:tau}. Equation
\eqref{eq:pyr-Zlim} then gives
\begin{equation}\label{eq:pyr-radial}
 -16D_{\rm p}\bigl(\mathscr Z_0\mathcal A_0\bigr)
    = D_{\rm p}(D_{\rm p}-1)(D_{\rm p}-2)\tau_{\rm G}.
\end{equation}
This recovers the radial third-order GWW action after an additional
Euler derivative. The full standard GWW system
\eqref{eq:gTriple}--\eqref{eq:gthree} follows on the limiting
quartic, where both independent differences have zero polynomial
remainder (Appendix~\ref{app:verify}). It is not a componentwise
limit of \eqref{eq:pyr-Z}. For example, the QCD exchange $a\leftrightarrow b$
induces $(v,H)\mapsto(-v,-H)$ in the chosen scaling, not an
exchange of the GWW charge and source. This is consistent with
the transverse rescaling in \eqref{eq:transverse} and explains
why the shared abstract $S_4$ does not identify the two triples
in this degeneration.

\section{Differential order and ordinary equations in a finite algebra}
\label{sec:ordinary}

\subsection{Algebraic degree, PDE order, and fixed-parameter ODE order}

The quartic and sextic count algebraic branches at generic parameters.
Their degrees do not specify the order of every differential equation
satisfied by a particular generating function. A mixed PDE may use
parameter derivatives that cannot be eliminated without increasing the
order in the remaining mass variable. The QCD equation
\eqref{eq:qInvariantPDE}, for example, has order two in
$(r,s,t)$, whereas an ordinary equation at fixed $r,s$ can have higher
order.

There are two further distinctions. First, an invertible change of
independent coordinates and an invertible affine change of dependent
variable preserve the order of a specified differential operator.
Finding an additional second-order equation for GWW does not turn its
known third-order equation into a second-order one by a coordinate
change. Second, replacing a potential by its derivative is not an
invertible point transformation. It loses integration data, and the
order of the corresponding equation can decrease by one. This is what
occurs in passing from \eqref{eq:qSinvariant} for the potential to
\eqref{eq:qInvariantPDE} for its normalized mass derivative.

The coefficient recursions illustrate the same distinction in discrete
form. GWW has a third-order operator acting explicitly on the preceding
coefficient. QCD has second-order parameter derivatives acting on both
adjacent coefficients. The latter is an implicit relation and is not a
lower-order explicit raising formula. On special parameter slices the
algebraic curve and its differential equations can degenerate further;
such reductions do not determine the generic order.

\subsection{Differentiation in the quotient algebra}

A uniform construction of ordinary equations is available directly from
an algebraic polynomial. Let $P(t,w)$ be monic and separable of degree
$d$ in $w$, and let $\mathbb K$ be the rational-function field in $t$ and the
fixed external parameters. In the finite algebra
\begin{equation}
 \mathcal A=\mathbb K[w]/(P),
\end{equation}
differentiation along a local root is represented by
\begin{equation}\label{eq:connection}
 \nabla_t f=\left[\partial_t f-\partial_wf\,P_tP_w^{-1}\right]\bmod P.
\end{equation}
The inverse of $P_w$ is taken modulo $P$. It exists on the regular
part of the finite covering, where the polynomial is separable. All
successive derivatives of an algebraic function can therefore be
represented in a fixed finite-dimensional space, using rational
coefficients and polynomial reduction.

The trace in this algebra is computable without writing its roots:
\begin{equation}\label{eq:trace}
 \Tr_{\mathcal A/\mathbb K}(f)=[w^{d-1}]\remP_P(fP_w).
\end{equation}
If a centered element $V$ has zero trace, all its derivatives have zero
trace as well. They lie in a space of dimension at most $d-1$. Set
$v_j=\nabla_t^jV$ and form
\begin{equation}
 H_{ij}=\Tr(v_iv_j),\qquad b_i=\Tr(v_iv_{d-1}),\qquad 0\leq i,j\leq d-2.
\end{equation}
On a chart where $\det H\neq0$, the coefficients of an ordinary
annihilator are
\begin{equation}\label{eq:traceODE}
 \left[\partial_t^{d-1}-\sum_{j=0}^{d-2}\gamma_j\partial_t^j\right]V=0,
 \qquad \gamma_j=\frac{\det H[j\leftarrow b]}{\det H}.
\end{equation}
This is a closed rational formula even when expanded coefficients are
large. The determinant also identifies where the chosen cyclic basis
fails; some resulting poles can be apparent rather than physical
singularities. Establishing the minimal possible order requires a
separate rank argument, not just the degree count.

For the depressed GWW quartic, $\tau$ has zero trace, and the
construction gives an equation of order at most three at fixed charge.
For QCD the coefficient of $W^5$ is $1/2$, so the six roots sum to
$-1/2$. Hence
\begin{equation}
 V_{\mathrm Q}=W+\frac1{12},\qquad\Tr V_{\mathrm Q}=0,
\end{equation}
and the centered mass-dependent algebraic function has an ordinary
annihilator of order at most five. This order is compatible with the
second-order mixed PDE because the latter also differentiates the
external invariants.

The relation to the uncentered mass derivative can be stated
explicitly. At fixed $k,u$, let
\begin{equation}
 L_5=\partial_t^5-\sum_{j=0}^4\gamma_j\partial_t^j,
 \qquad\mathcal D_t=\partial_t+\frac1{2t}.
\end{equation}
Using $z^2=t$ and \eqref{eq:qW},
\begin{equation}
 R=\frac{E_0}{kz}V_{\mathrm Q}+\frac{6k^2-E_0}{12kz}.
\end{equation}
The extra elementary term can be eliminated by an additional first-order
factor. On a generic chart with $\gamma_0\neq0$, this gives
\begin{equation}\label{eq:Rordinary}
 \left(\mathcal D_t-\frac{\partial_t\gamma_0}{\gamma_0}\right)
 \left(\mathcal D_t^5-\sum_{j=0}^4\gamma_j\mathcal D_t^j\right)R=0.
\end{equation}
Products here mean composition of noncommuting differential operators:
the outer derivative also differentiates inner coefficients. Equation
\eqref{eq:Rordinary} is an ordinary sixth-order representation at
fixed parameters. It does not replace or contradict the third-order
mixed equation for $S$.

\section{Structural comparison and scope of the formulation}
\label{sec:discussion}

The two models share a projective parameter geometry, but their
algebraic and differential realizations retain distinct features. The
GWW invariant quartic requires an orientation covariant unless the
algebraic variable is multiplied by the same orientation factor. The
QCD sextic is symmetric in the squared homogeneous coordinates from the
outset. In GWW the source is part of the projective triple; in QCD the
mass appears as a separate invariant coordinate. These differences
explain why the common signed-permutation group does not make the two
operator systems identical at generic parameters.

The comparison can be summarized as follows.
\begin{center}\small
\renewcommand{\arraystretch}{1.3}
\begin{tabular}{R{.20\linewidth}R{.35\linewidth}R{.35\linewidth}}
\toprule
Structure&GWW&One-dimensional QCD\\\midrule
Projective triple&$[u:h:i]$&$[k:k+1:ku]$\\
Parameter group&Signed permutations, $S_4$&Signed permutations, $S_4$\\
Operator labels&Quotient action $S_3$&Quotient action $S_3$\\
Algebraic degree&Four&Six\\
Source or mass&Contained in the triple&Additional coordinate $t$\\
Homogeneous operators&$a_i\partial_{a_i}^3$ on $\tau$&$\mathscr K_i$ on $Y_{\mathrm Q}$; $\mathscr T_i$ on $\Phi$\\
Triple relation&Pairwise equality&Weighted cyclic equation\\
Coefficient recursion&Explicit third-derivative raising&Two-step row relation and explicit column reconstruction\\
Model connection&Many-flavor limit and first column&Sextic pyramid with GWW and Narayana boundaries\\
Tree interpretation&Description-tree refinement (conjectural)&Decorated bipartite trees on a collapsed slice (exact)\\\bottomrule
\end{tabular}
\end{center}

The cyclic QCD equation explains how an alternating complete
operator can coexist with permuted components. The counterexample
\eqref{eq:qTripleCounterexample} concerns the original actions
on the shifted free energy; the divided differences
\eqref{eq:pyr-Z} have equal actions precisely because they express
one weighted equation. Their common GWW boundary operator is
given by \eqref{eq:pyr-Zlim}. The full third-order GWW system
comes from the limiting quartic, with the role of the first
transverse coefficient made explicit by \eqref{eq:pyr-boundary-PDE}.

The row recurrence \eqref{eq:pyr-row-rec} is a differential
constraint on two rows, not a one-step raising operator.
Nevertheless, \eqref{eq:pyr-column-rec} gives an explicit
reconstruction in the formal class, using GWW data and a
diagonal Euler inverse. This statement is separate from
generation by the original small-$h$ rational ansatz
in Appendix~\ref{app:original-qcd-rec}, for which an all-order
uniqueness theorem in that ansatz would be an additional result.

The combinatorial comparison likewise has a precise scope.
Equation~\eqref{eq:pyr-GWW} embeds the GWW series as one boundary of
the QCD pyramid, and \eqref{eq:pyr-Nar-invert} gives the Narayana
triangle on the other. The exact identity \eqref{eq:pyr-H-polar-identity}
recovers the critical GWW series at the Narayana pole. The positivity
certificates in \eqref{eq:pyr-H-positive} apply through the sixth
deformation order in every row. Equation~\eqref{eq:pyr-product} counts a
different, formally collapsed slice by two vertex colors. A tree statistic retaining the
full $x$ dependence, and a direct relation between the two tree
families, remain to be established. Color exchange of the collapsed
array is not by itself an identification with the projective $S_4$
action on the original physical parameters.

Finally, the projective symmetries describe analytic algebraic branches.
They do not by themselves select the dominant physical saddle after a
charge extremization or a sector sum. The real GWW parameterization
provides a useful branch prescription within its one-cut family, while
the QCD small-$h$ boundary condition identifies a fixed-sector germ.
Global continuation across physical phase boundaries, complex-source
cancellations, and alternative operator decompositions with a more
direct contraction to GWW remain separate questions. These questions can be addressed
using the algebraic curves and differential systems without changing
the normalization dictionary developed above.

\appendix
\section{Exact algebraic verification of the differential identities}
\label{app:verify}

The differential identities can be checked without selecting radical
expressions for individual roots. Let $P(w,\boldsymbol\xi)=0$ be a
monic algebraic relation and assume $P_w\neq0$ on a regular branch.
Implicit differentiation gives
\begin{equation}
 w_i=-\frac{P_i}{P_w},\qquad
 w_{ij}=\frac{J_{ij}}{P_w^3},
\end{equation}
with
\begin{equation}\label{eq:Jij}
 J_{ij}=-P_{ij}P_w^2+(P_{iw}P_j+P_{jw}P_i)P_w
 -P_{ww}P_iP_j.
\end{equation}
Multiplying a proposed second-order PDE by $P_w^3$ reduces its
verification to a polynomial identity modulo $P$.

For the QCD equation \eqref{eq:qInvariantPDE}, the exact remainder is
\begin{align}\label{eq:qremainder}
0=\remP_{P_6}\bigl[&2r(4r-1)J_{rr}+4s(6r-1)J_{rs}
 +18s^2J_{ss}+2t(t-1)J_{rt}\notag\\
&-(4r-t)P_rP_w^2-12sP_sP_w^2\bigr].
\end{align}
Every coefficient vanishes in rational polynomial arithmetic. This
proves the relation for every regular local root of the sextic. The
same procedure verifies \eqref{eq:gSecond} against the canonical
quartic. At a ramification point, the displayed implicit-derivative
fractions require a local continuation analysis, even though the
polynomial identities remain meaningful.

For a third derivative along a parameter derivation $\delta$, such as
$\partial_u$, $\partial_h$, or $u\partial_u+h\partial_h$, define
\begin{align}
 J_\delta={}&-(\delta^2P)P_w^2
 +2(\delta P_w)(\delta P)P_w-P_{ww}(\delta P)^2,\\
 N_\delta={}&[(\delta J_\delta)P_w-(J_\delta)_w\delta P]P_w\notag\\
 &-3J_\delta[(\delta P_w)P_w-P_{ww}\delta P].
\end{align}
Then $\delta^2w=J_\delta/P_w^3$ and
$\delta^3w=N_\delta/P_w^5$. For GWW, the two independent equations
in \eqref{eq:gthree} reduce to
\begin{equation}
 \remP_{P_4}(hN_{\partial_h}-uN_{\partial_u})=0
\end{equation}
and
\begin{equation}
 \remP_{P_4}\bigl[N_D-3J_DP_w^2-2(DP)P_w^4-uN_{\partial_u}\bigr]=0.
\end{equation}
Both remainders vanish exactly. The second formula accounts for the
lower powers of $D$ in $\Delta_0=D^3-3D^2+2D$.

The distinction between these all-parameter verifications and a series
check is useful in practice. A finite list of coefficients can reveal
or refute a proposed relation, but it does not prove the identity on
the complete curve. Conversely, an exact remainder proves the
algebraic identity but does not select its physical branch. These are
complementary checks with different purposes.

\section{Derivation of the cyclic QCD decomposition}
\label{app:cyclic}

The cyclic construction can be made explicit on the homogeneous chart
used in \eqref{eq:qOperatorIdentity}. For a degree-one function
$Y(A_1,A_2,A_3;t)$ with $A_1=1$, write $B=A_2$ and $C=A_3$.
Homogeneity implies
\begin{equation}
 Y_{A_1}=Y-BY_B-CY_C,\qquad
 Y_{A_1A_1}=B^2Y_{BB}+2BCY_{BC}+C^2Y_{CC}.
\end{equation}
Using these formulas in \eqref{eq:qKcyclic} gives the differential
expression
\begin{align}\label{eq:expandedCyclic}
\mathcal CY={}&B^2(B-1)Y_{BB}+2BC(B-C)Y_{BC}-C^2(C-1)Y_{CC}\notag\\
&+\eta\bigl[B(B-2C+1)Y_{Bt}
 +C(2B-C-1)Y_{Ct}-(B-C)Y_t\bigr]\notag\\
&-\frac t2\bigl[B(B-2C+1)Y_B
 +C(2B-C-1)Y_C-(B-C)Y\bigr].
\end{align}
The form of the coefficients follows from the same cyclic differences
that multiply the three Euler operators. Substitution of
$r=(B+C+BC)/(1+B+C)^2$, $s=BC/(1+B+C)^3$, and $W=Y/(1+B+C)$
into \eqref{eq:qInvariantPDE} yields
\eqref{eq:qOperatorIdentity} term by term. This supplies a direct
operator-level derivation of the triple from the invariant PDE.

The logarithmic leading term also admits a simple homogeneous check.
In the physical chart it has coefficient
$L=ku-k(k+1)$. Its homogeneous lift is
\begin{equation}
 L(a,b,c)=bc-ac-ab.
\end{equation}
For each squared-coordinate Euler operator,
$D_i(D_i-1/2)L=0$, because every monomial contains each unsquared
coordinate at most once. Applying \eqref{eq:qTx} to
$(L/2)\log x$ gives
\begin{equation}
 \mathscr T_i^{(x)}\left[\frac L2\log x\right]
 =\frac{1+x}{x}D_i(D_i-1/2)L=0.
\end{equation}
Thus the logarithmic leading term is annihilated by each component
separately. The coefficient recursion \eqref{eq:qphiRec} follows
from the regular part alone, without an omitted inhomogeneous
contribution from the leading logarithm.

Finally, inserting $\sum_{n\geq1}\phi_nx^n$ into
\eqref{eq:qPhiPDE} gives the coefficient of $x^n$ as
\begin{equation}
 (n+1)[2U-(2n+3)V]\phi_{n+1}
 +n[2U+(2n-1)V]\phi_n.
\end{equation}
The $x^0$ coefficient gives $(2U-3V)\phi_1=0$. This derivation
also explains why derivatives of the new coefficient are an intrinsic
part of the recurrence, rather than an artifact of notation.

\section{Recurrences in the original QCD parameters}
\label{app:original-qcd-rec}

These formulas use the expansion at fixed $(k,u)$ in powers of
$x=h^2$. They give complementary differential constraints to
the resummed small-charge recurrence in Section~\ref{sec:pyramid-rows}.
The normalization dictionary is useful when computing the original
free-energy coefficients or checking the action of projective maps.

\subsection{The small-\texorpdfstring{$h$}{h} series and an auxiliary generating function}

For comparison with the pyramid expansion, use the physical variables
$k,u$. Write
\begin{equation}\label{eq:qSseries}
 S=S_{\mathrm{leading}}+\sum_{n\geq1}C_n(k,u)x^n,
 \qquad\partial_hS_{\mathrm{leading}}=\frac{u-1}{h}.
\end{equation}
An auxiliary generating function $F=\sum_{n\geq0}f_nx^n$ is defined
by
\begin{equation}\label{eq:qF}
 \G_{\mathrm Q}=\frac{ku}{2}
 +\frac{k^2x-k(k+1)xF}{1-x}.
\end{equation}
It is related to the original action coefficients by
\begin{equation}\label{eq:qCfromf}
 C_n=\frac{k-(k+1)\sum_{j=0}^{n-1}f_j}{n}.
\end{equation}
Conversely,
\begin{equation}
 f_0=\frac{k-C_1}{k+1},\qquad
 f_n=\frac{nC_n-(n+1)C_{n+1}}{k+1}\quad(n\geq1).
\end{equation}
The distinction between $f_n$ and $C_n$ is substantive: an operator
acting on one of these sequences must be conjugated or combined with
the coefficient relation before it is applied to the other.

The first auxiliary coefficients are
\begin{align}\label{eq:qfirstf}
f_0={}&\frac{ku}{1+ku},\\
f_1={}&\frac{k^2u(1-u)[1+k(1+u)]}{(1+ku)^4},\\
f_2={}&\frac{k^2u(1-u)[1+k(1+u)]}{(1+ku)^7}\notag\\
&\times\bigl[(1+ku)(1+k^2u^2)+2k(k+1)(1-ku)\bigr].
\end{align}
In particular,
\begin{equation}\label{eq:qC1}
 C_1=\frac{k(1-u)}{1+ku}.
\end{equation}
As a normalization check, at $k=1,u=1/2$ the first five action
coefficients are
\begin{equation}\label{eq:qCsample}
 (C_1,C_2,C_3,C_4,C_5)=
 \left(\frac13,\frac7{162},-\frac{431}{6561},
 -\frac{6739}{78732},-\frac{40159}{7971615}\right).
\end{equation}
At saturation $u=1$, all these regular coefficients vanish. At $u=0$,
the weak branch has $C_n=k/n$, giving
$\Psi_{\mathrm Q}=-k^2\log(1-h^2)$ with $\Psi_{\mathrm Q}(0)=0$.
This latter expression describes a local weak small-$h$ branch and should
not be extrapolated as the physical answer at arbitrary $h$ without a
separate branch analysis.

\subsection{The implicit differential recursion}

Introduce the coefficient polynomials
\begin{equation}
 a_0=2k^2(1-u^2)+2k+1,\qquad
 b_0=k^2(1+u^2)+k,\qquad c_0=k^2(u^2-1)-4k-2,
\end{equation}
and the second-order operator
\begin{align}\label{eq:qL}
L={}&k^2u^2(u^2-1)\partial_u^2
 +2k(k+1)u[(k+1)^2-k^2u^2]\partial_u\partial_k\notag\\
 &-k^2(k+1)^2(2k+1)\partial_k^2.
\end{align}
The two operators needed for the recursion are
\begin{align}\label{eq:qMP}
M={}&L-2k(k+1)b_0\partial_k+2uc_0\partial_u+2(2k+1),\\
P={}&2k(k+1)a_0\partial_k+2uc_0\partial_u+2(2k+1).
\end{align}
Terms without derivatives denote multiplication. Although $M$ and $P$
enter symmetrically in the notation, their differential orders differ:
$M$ is second order and $P$ is first order.

The exact adjacent-coefficient relation is
\begin{equation}\label{eq:qfRec}
 (M+nP)f_n=(nP-M)f_{n-1},\qquad n\geq1.
\end{equation}
This is a differential constraint on both the new and the previous
coefficient. It can be used effectively once the rational class of
$f_n$ is specified, but it is not an explicit second-order replacement
for the GWW raising operator. In particular, the left-hand derivatives
cannot be discarded when testing the recurrence.

A finite ansatz compatible with the first coefficients is
\begin{equation}\label{eq:qansatz}
 f_n=\frac{k^2u(1-u)[1+k(1+u)]}{(1+ku)^{3n+1}}
 \sum_{a=0}^{n-1}\sum_{b=0}^{3(n-1)-2a}
 c_{ab}[k(k+1)]^a(ku)^b,\qquad n\geq1.
\end{equation}
Substituting this ansatz into \eqref{eq:qfRec} gives a finite linear
system for the coefficients $c_{ab}$. The system has a unique solution
in this class for the first five nontrivial coefficients that have been
checked. This supplies a concrete generation procedure; a proof of
uniqueness for every $n$ would be an additional result.

Using \eqref{eq:qCfromf}, the recurrence can also be written directly
for the original free-energy coefficients:
\begin{equation}\label{eq:qCRec}
 (n+1)(M+nP)\left[\frac{C_{n+1}}{k+1}\right]
 =n[(n+1)P-M]\left[\frac{C_n}{k+1}\right].
\end{equation}
The denominator belongs inside the argument of the operator. Since the
operators differentiate in $k$, moving $1/(k+1)$ outside them changes
the recurrence. This is a useful practical distinction when checking
individual coefficients in a symbolic calculation.

\subsection{The equation for the unshifted free energy}

The coefficient relation can be summed into a PDE for the complete
free energy. Define
\begin{equation}
 \mathsf A=(k+1)M(k+1)^{-1},\qquad
 \mathsf B=(k+1)P(k+1)^{-1},\qquad
 D_0=(k+1)^2-k^2u^2.
\end{equation}
Explicitly,
\begin{align}\label{eq:qAB}
\mathsf A={}&L+2k^3(k+1)(1-u^2)\partial_k
 +2u(c_0-kD_0)\partial_u\notag\\
 &+2[k^3(u^2-1)+2k+1],\\
\mathsf B={}&2k(k+1)a_0\partial_k+2uc_0\partial_u
 +2(2k+1-ka_0).
\end{align}
The full equation, including the logarithmic leading term, is
\begin{align}\label{eq:qSPDE}
 \mathcal ES:={}&h(1-h^2)\mathsf B[S_{hh}]\notag\\
 &+\bigl[2(1+h^2)\mathsf A-(1+3h^2)\mathsf B\bigr][S_h]
 =-\frac{4k(1-h^2)}hD_0=:\mathcal R.
\end{align}
The first term contains one parameter derivative and two mass
derivatives. The second contains up to two parameter derivatives and
one mass derivative. The total order is therefore three, even though
$S_{hh}$ appears multiplied by a differential operator. Replacing the
coefficient index by a source derivative does not produce a fourth
order here because the coefficient of the higher source derivative is
only first order in the external parameters.

The operators in \eqref{eq:qSPDE} are not literally invariant scalar
operators on $S$. If $T^*$ denotes parameter pullback, the exact laws
are
\begin{equation}\label{eq:qABcov}
 T^*\mathsf A(T^*)^{-1}=\rho_T\alpha_T\mathsf A\alpha_T^{-1},
 \qquad
 T^*\mathsf B(T^*)^{-1}=\rho_T\alpha_T\mathsf B\alpha_T^{-1},
\end{equation}
with
\begin{equation}
 \rho_{ab}=-1,\qquad
 \rho_{bc}=-\frac1{k^2(u-1)^2},\qquad
 \rho_{ac}=-\frac1{(k+1-ku)^2}.
\end{equation}
Consequently
\begin{equation}
 \mathcal E_T=\rho_T\alpha_T\mathcal E\alpha_T^{-1},\qquad
 T^*\mathcal R=\rho_T\alpha_T\mathcal R+\mathcal E_T[\beta_T\ell].
\end{equation}
The source term and the logarithmic part of the free-energy
transformation are thus tied to the operator covariance. For example,
$T_{ab}^*\mathsf B(T_{ab}^*)^{-1}=-\mathsf B+2a_0$, which
exhibits its literal noninvariance. After the weight of $S$ and the
affine logarithmic term are included, the transformed equation is
equivalent to the original one.

This covariance concerns a specific complete operator and its chosen
two-component representation. It does not rule out a different
representation in which a triple of components is permuted. The
homogeneous potential \eqref{eq:qPhiLift} makes precisely that
representation possible.

\subsection{Physical-coordinate realization and the original action coefficients}

The homogeneous construction can be used without differentiating in
three new independent parameters. To derive its physical realization,
write $d_0=b-a$, $k=a/d_0$, $u=c/a$, and
$\Phi(a,b,c)=d_0^2\Phi(k,u)$. Applying
$D_i=(a_i/2)\partial_{a_i}$ and then setting $d_0=1$ gives
\begin{equation}\label{eq:qDi}
\begin{aligned}
D_1&=\frac{k(k+1)}2\partial_k-\frac u2\partial_u-k,\\
D_2&=-\frac{k(k+1)}2\partial_k+(k+1),\\
D_3&=\frac u2\partial_u.
\end{aligned}
\end{equation}
The zeroth-order terms are the contributions of the homogeneous
weight. They ensure $D_1+D_2+D_3=1$ and the commutativity of the
physical-coordinate representation. Removing them would no longer
represent the same Euler operators on the degree-one functions.

The cyclic weights become
\begin{equation}\label{eq:qweights}
 w_1=(k+1)^2-k^2u^2=D_0,\qquad
 w_2=k^2(u^2-1),\qquad w_3=-(2k+1).
\end{equation}
They multiply from the left in \eqref{eq:qTcyclic}. The parameter
derivatives inside each $\mathscr T_i$ do not act on the external
weight $w_i$. Define
\begin{equation}\label{eq:qUV}
 U=\sum_{i=1}^3w_iD_i^2,\qquad V=\sum_{i=1}^3w_iD_i.
\end{equation}
Here $D_i^2$ denotes operator composition, including differentiation of
its internal coefficients. The cyclic equation has the compact form
\begin{equation}\label{eq:qPhiPDE}
 2(1+x)U[\Phi_x]+(x-3)V[\Phi_x]
 -2x(1-x)V[\Phi_{xx}]=0.
\end{equation}
Substitution of $\Phi=kS-k^2\ell$ shows that its left-hand side is
exactly $-k/2$ times the residual of \eqref{eq:qSPDE} expressed in
$x$, namely the residual in $h$ divided by $4h$. This checks the
source term and logarithmic shift together with the principal
higher-derivative terms.

The relation to the original small-source coefficients is particularly
simple:
\begin{equation}\label{eq:qphiseries}
 \Phi=\frac{ku-k(k+1)}2\log x+\Phi_0(k,u)
 +\sum_{n\geq1}\phi_n(k,u)x^n,
 \qquad \phi_n=kC_n-\frac{k^2}{n}.
\end{equation}
The leading logarithm separately satisfies \eqref{eq:qPhiPDE}; the
function $\Phi_0$ is annihilated by the mass derivative. Thus the
regular coefficients obey the homogeneous differential recursion
\begin{equation}\label{eq:qphiRec}
 (n+1)[2U-(2n+3)V]\phi_{n+1}
 =-n[2U+(2n-1)V]\phi_n,\qquad n\geq1.
\end{equation}
The initial coefficient and its equation are
\begin{equation}
 \phi_1=-\frac{k^2(k+1)u}{1+ku},\qquad (2U-3V)\phi_1=0.
\end{equation}
The next one is
\begin{equation}
 \phi_2=-\frac{k^2u(k+1)[k^3u^3+2k^2u^2+k^2+2ku+k+1]}
 {2(1+ku)^4}.
\end{equation}
At $k=1,u=1/2$,
\begin{equation}
 (\phi_1,\phi_2,\phi_3)=
 \left(-\frac23,-\frac{37}{81},-\frac{2618}{6561}\right).
\end{equation}
After the parameter derivatives are evaluated, the two sides of
\eqref{eq:qphiRec} equal $6$ at $n=1$ and $250/9$ at $n=2$.
Four consecutive transitions, through $n=4$, have been checked exactly
against the rational coefficients obtained independently from the
auxiliary recurrence.

The cyclic structure is genuinely different from the pairwise equality
in \eqref{eq:gTriple}. In the same example, the three actions at
$x\to0$ are
\begin{equation}\label{eq:qTripleCounterexample}
 (\mathscr T_1^{(x)}\Phi,\mathscr T_2^{(x)}\Phi,
 \mathscr T_3^{(x)}\Phi)=
 \left(\frac{10}{27},-\frac{14}{27},\frac{16}{27}\right),
\end{equation}
while $(w_1,w_2,w_3)=(15/4,-3/4,-3)$ gives a zero weighted sum.
Replacing the cyclic equation by pairwise equality would therefore
impose an incorrect additional condition. A different normalization or
additional operators would be needed to realize that stronger analogy.
The present formulation already supplies the permuted components and
the exact third-order equation, without assuming the stronger property.

The counterexample concerns these original components on the
shifted free energy. The divided differences in \eqref{eq:pyr-Z}
act on the regular pyramid derivative and re-express one cyclic
constraint; their equal actions do not contradict this test.

\section{Coefficient extraction and exact checks for the pyramid}
\label{app:pyramid-checks}

The coefficient form of \eqref{eq:pyr-column-rec}, useful when
working at fixed row $m$, is
\begin{align}\label{eq:pyr-cell}
 (j+1)t_{\rm p}E_m A_{m,j+1}
 ={}&[t_{\rm p}E_m(j-H_m)-2(H_m-2j)(H_m+1)]A_{m,j}
 \notag\\
 &+2E_{m-2}(E_{m-2}+2j+1)A_{m-2,j}\notag\\
 &+4E_{m-2}(H_{m-2}-2j+2)A_{m-2,j-1}\notag\\
 &+(4j-6m+4)E_{m-2}A_{m-2,j-2}.
\end{align}
Terms with a negative row or column index are zero. The diagonal
inverse of $E_m$ on $t_{\rm p}^b$ divides by $m-1+2b$.
This is the coefficient form of \eqref{eq:pyr-inverse}, so
it does not require solving a differential boundary-value problem
in $t_{\rm p}$.

For a direct operator check of \eqref{eq:pyr-cyclic-identity},
restrict \eqref{eq:pyr-dX} to the coefficient of $\zeta^{m+1}$
and write $\mathscr J_{i,m}$ for the resulting operator.
Let $f=\delta(1-x)/[8(1+x)]$. Then
\begin{align}
 \tfrac14(\delta+1)^2\mathscr J_{1,m}
 -\tfrac14(\delta-1)^2\mathscr J_{2,m}-\delta\mathscr J_{3,m}
 &=-f\mathscr L_m,\notag\\
 (1-x)^2(\mathscr J_{2,m}-\mathscr J_{1,m})
 &=2 f E_m\mathscr M_m .
\end{align}
The first line collects the charge-independent part of the weights;
the second multiplies the contribution shifted by $\zeta^2$.
Composition in these identities differentiates the internal
coefficient $\delta$. The symbolic operator check uses an arbitrary
test function; the additional checks on ten explicit rows are
not needed for the identity.

To obtain the second-column relation independently of the
coefficient table, put
$p_0=\widetilde P|_{x=0}/t_{\rm p}^2$ and
$p_1=\partial_x\widetilde P|_{x=0}/t_{\rm p}^2$.
At $a=\mathcal A_0$ the first transverse coefficient is
$\mathcal A_1=-p_1/p_{0,a}$. Define the derivative along the
quartic by
$\nabla_i=\partial_i-(p_{0,i}/p_{0,a})\partial_a$ for
$i=\zeta,t_{\rm p}$. Clearing denominators in
\begin{equation}
 \nabla_\zeta\left(-\frac{p_1}{p_{0,a}}\right)
        -6t_{\rm p}\zeta\nabla_{t_{\rm p}}^{\,2}a
\end{equation}
gives a numerator of zero polynomial remainder modulo $p_0$.
This proves the relation at every regular point of the
algebraic expansion, and hence as a formal identity for the
selected germ. The companion scripts construct the cleared
curve, perform this reduction, and check the leading factor
used in the Narayana derivation.

\section{The collapsed QCD slice: derivation and enumeration}
\label{app:qcd-trees}

We give an algebraic proof connecting the tree construction to the
QCD sextic, rather than infer that connection from a finite table.
Use the substitution \eqref{eq:pyr-curve-sub} in the cleared
polynomial \eqref{eq:rawQ}, with $a=\mathcal A_{\rm p}(\zeta,4\tau,x)$.
After this substitution $\mathcal Q$ has a factor $(1-x)^{10}$.
Dividing by this factor \emph{before} taking $x\to1$ gives exactly
the following quintic for $g=\mathcal T_{\mathrm Q}$, with $a=g$.
Put $\sigma=\zeta+\tau$ and $\pi=\zeta\tau$. Then
\begin{equation}\label{eq:pyr-quintic}
 F(g;\zeta,\tau)=16g^5+c_4g^4+c_3g^3+c_2g^2+c_1g+c_0=0,
\end{equation}
with
\begin{align}
 c_4={}&-32\pi+16\sigma^2-40\sigma+1,\notag\\
 c_3={}&2(8\pi^2+32\pi\sigma+8\pi-16\sigma^3
                         +16\sigma^2-\sigma),\notag\\
 c_2={}&-24\pi^2\sigma-62\pi^2-16\pi\sigma^2-24\pi\sigma
                      +\pi+16\sigma^4-8\sigma^3+\sigma^2,\notag\\
 c_1={}&\pi(72\pi^2-24\pi\sigma^2+62\pi\sigma-2\pi
                           -16\sigma^3+8\sigma^2-\sigma),\notag\\
 c_0={}&-\pi^2(27\pi^2+36\pi\sigma-\pi
                           -16\sigma^3+8\sigma^2-\sigma).
\label{eq:pyr-quintic-coeffs}
\end{align}
The coefficient germ is $g=\zeta\tau+O_{\rm tot}(3)$.
The polynomial is symmetric in $\zeta,\tau$, but its derivative
with respect to $g$ is not a unit at the origin, so the ordinary
implicit-function theorem does not establish branch uniqueness there.
Instead assign weights $2,1,1$ to $g,\zeta,\tau$. The lowest
weighted part of $F$ is
\begin{equation}
 \zeta\tau(\zeta+\tau)(\zeta\tau-g).
\end{equation}
For two putative roots with the indicated germ, the divided difference
of $F$ has leading term $-\zeta\tau(\zeta+\tau)\neq0$.
Since the formal power-series ring is an integral domain, the two
roots coincide. Thus the germ, whenever constructed, is unique.

To check the tree expression exactly, put
\begin{equation}\label{eq:pyr-rational-param}
 U=\frac{1-r_0^2}{4},\quad V=\frac{1-s_0^2}{4},\quad
 \zeta=Us_0,\quad\tau=Vr_0,
 \qquad r_0=s_0=1\text{ at the origin}.
\end{equation}
Equations \eqref{eq:pyr-planted} hold, and the right-hand side
$g_{\rm tree}$ of \eqref{eq:pyr-tree-GF} becomes
\begin{equation}
 16g_{\rm tree}=(r_0-1)(s_0-1)(3r_0s_0+r_0+s_0-1).
\end{equation}
Substitution into \eqref{eq:pyr-quintic} gives the zero polynomial
in $r_0,s_0$, with no series truncation. Its germ is
$\zeta\tau+O_{\rm tot}(3)$, establishing the identity with the
collapsed QCD branch by uniqueness.

For completeness, coefficient extraction gives the product formula
without eliminating $U,V$. Every tree counted by
$\zeta^M\tau^N$ has $M+N-1$ edges. Edge marking therefore yields
\begin{equation}\label{eq:pyr-edge-marking}
 (\zeta\partial_\zeta+\tau\partial_\tau-1)
                 \mathcal T_{\mathrm Q}=UV.
\end{equation}
The multivariate Lagrange--Good formula applied to
\eqref{eq:pyr-planted} reads
\begin{align}\label{eq:pyr-Good}
 [\zeta^M\tau^N]UV
 ={}&[u_0^{M-1}v_0^{N-1}]
 (1-4v_0)^{-M/2}(1-4u_0)^{-N/2}\notag\\
 &\hspace{7mm}\times
 \left(1-\frac{4u_0v_0}{(1-4u_0)(1-4v_0)}\right)\\
 ={}&\frac{4^{M+N-2}(N/2)_{M-1}(M/2)_{N-1}}
          {(M-1)!(N-1)!}
 \left(1-\frac{(M-1)(N-1)}{MN}\right).\notag
\end{align}
Combining \eqref{eq:pyr-edge-marking} and \eqref{eq:pyr-Good},
then setting $M=m+1$, $N=a+1$, proves
\eqref{eq:pyr-product}. The locally marked corners also rule out
nontrivial automorphisms: a tree automorphism fixes its central vertex
or its central edge; the colors prevent exchanging the endpoints of
the latter, and a fixed vertex with a marked corner fixes all its
incident edges. This justifies ordinary, rather than
automorphism-weighted, counting in the dissymmetry identity.

\end{document}